\pdfoutput=1
\PassOptionsToPackage{unicode}{hyperref}
\PassOptionsToPackage{hyphens}{url}
\PassOptionsToPackage{dvipsnames,svgnames,x11names}{xcolor}
\PassOptionsToPackage{format=plain,
  labelfont={bf,sf,small,singlespacing},
  textfont={sf,small,singlespacing},
  justification=justified,
  margin=0.25in}{caption}%
\documentclass[
  11pt,
]{article}
\usepackage{xcolor}
\usepackage[lmargin=3cm,rmargin=3cm,tmargin=2.5cm,bmargin=2.5cm]{geometry}
\usepackage{amsmath,amssymb}
\usepackage{iftex}
\ifPDFTeX
  \usepackage[T1]{fontenc}
  \usepackage[utf8]{inputenc}
  \usepackage{textcomp} % provide euro and other symbols
    \usepackage{mathpazo}
  \usepackage{newpxtext}
      \usepackage[scaled]{helvet}
    \usepackage[scale=0.95,varl]{inconsolata}
\else % if luatex or xetex
          \usepackage{unicode-math} % this also loads fontspec
        \usepackage{amsthm}
    \usepackage{newtxmath}
    \usepackage{newtxtext}
        \usepackage[scaled]{helvet}
        \defaultfontfeatures{Scale=MatchLowercase}
    \defaultfontfeatures[\rmfamily]{Ligatures=TeX,Scale=1}
  \fi

\usepackage{sectsty}
\ifPDFTeX\else
\fi
\IfFileExists{upquote.sty}{\usepackage{upquote}}{}
\IfFileExists{microtype.sty}{% use microtype if available
  \usepackage[]{microtype}
  \UseMicrotypeSet[protrusion]{basicmath} % disable protrusion for tt fonts
}{}
\makeatletter
\@ifundefined{KOMAClassName}{% if non-KOMA class
  \IfFileExists{parskip.sty}{%
    \usepackage{parskip}
  }{% else
    \setlength{\parindent}{0pt}
    \setlength{\parskip}{6pt plus 2pt minus 1pt}}
}{% if KOMA class
  \KOMAoptions{parskip=half}}
\makeatother

\usepackage{longtable,booktabs,array}
\usepackage{calc} % for calculating minipage widths
\usepackage{etoolbox}
\makeatletter
\patchcmd\longtable{\par}{\if@noskipsec\mbox{}\fi\par}{}{}
\makeatother
\IfFileExists{footnotehyper.sty}{\usepackage{footnotehyper}}{\usepackage{footnote}}
\makesavenoteenv{longtable}
\usepackage{graphicx}
\makeatletter
\newsavebox\pandoc@box
\newcommand*\pandocbounded[1]{% scales image to fit in text height/width
  \sbox\pandoc@box{#1}%
  \Gscale@div\@tempa{\textheight}{\dimexpr\ht\pandoc@box+\dp\pandoc@box\relax}%
  \Gscale@div\@tempb{\linewidth}{\wd\pandoc@box}%
  \ifdim\@tempb\p@<\@tempa\p@\let\@tempa\@tempb\fi% select the smaller of both
  \ifdim\@tempa\p@<\p@\scalebox{\@tempa}{\usebox\pandoc@box}%
  \else\usebox{\pandoc@box}%
  \fi%
}
\def\fps@figure{htbp}
\makeatother

\ifLuaTeX
  \usepackage{luacolor}
  \usepackage[soul]{lua-ul}
\else
  \usepackage{soul}
\fi

\providecommand{\tightlist}{%
  \setlength{\itemsep}{0pt}\setlength{\parskip}{0pt}}

\usepackage{xcolor}
\colorlet{RevisionColor}{black}
\newenvironment{rrchange}{\begingroup\color{RevisionColor}}{\endgroup}

\newcommand{\qand}{\quad\text{and}\quad}
\newcommand{\eps}{\varepsilon}
\newcommand{\iid}{\stackrel{\text{iid}}{\sim}}
\DeclareMathOperator{\Norm}{\mathcal{N}}

\renewcommand{\bar}{\overline}
\DeclareMathOperator{\E}{\mathbb{E}}

\newcommand{\vb}[1]{\mathbf{#1}}
\newcommand{\vbg}[1]{\boldsymbol{#1}}

\usepackage[hang, bottom]{footmisc} % bottom forces fn at end even if there is whitespace
\allowdisplaybreaks

\usepackage[authordate, backend=biber]{biblatex-chicago}
\usepackage{booktabs}
\usepackage{caption}
\usepackage{longtable}
\usepackage{colortbl}
\usepackage{array}
\usepackage{anyfontsize}
\usepackage{multirow}
\makeatletter
\@ifpackageloaded{caption}{}{\usepackage{caption}}
\AtBeginDocument{%
\ifdefined\contentsname
  \renewcommand*\contentsname{Table of contents}
\else
  \newcommand\contentsname{Table of contents}
\fi
\ifdefined\listfigurename
  \renewcommand*\listfigurename{List of Figures}
\else
  \newcommand\listfigurename{List of Figures}
\fi
\ifdefined\listtablename
  \renewcommand*\listtablename{List of Tables}
\else
  \newcommand\listtablename{List of Tables}
\fi
\ifdefined\figurename
  \renewcommand*\figurename{Figure}
\else
  \newcommand\figurename{Figure}
\fi
\ifdefined\tablename
  \renewcommand*\tablename{Table}
\else
  \newcommand\tablename{Table}
\fi
}
\@ifpackageloaded{float}{}{\usepackage{float}}
\floatstyle{ruled}
\@ifundefined{c@chapter}{\newfloat{codelisting}{h}{lop}}{\newfloat{codelisting}{h}{lop}[chapter]}
\floatname{codelisting}{Listing}

\usepackage{amsthm}
\theoremstyle{plain}
\newtheorem{lemma}{Lemma}[section]
\theoremstyle{plain}
\newtheorem{proposition}{Proposition}[section]
\theoremstyle{remark}
\AtBeginDocument{}

\makeatother
\makeatletter
\@ifpackageloaded{caption}{}{\usepackage{caption}}
\@ifpackageloaded{subcaption}{}{\usepackage{subcaption}}
\makeatother
\usepackage{bookmark}
\IfFileExists{xurl.sty}{\usepackage{xurl}}{} % add URL line breaks if available
\hypersetup{
  pdftitle={Confounders and Linearity in Ecological Inference},
  pdfauthor={Shiro Kuriwaki; Cory McCartan},
  colorlinks=true,
  linkcolor={black},
  filecolor={Maroon},
  citecolor={VioletRed4},
  urlcolor={DodgerBlue4},
  pdfcreator={LaTeX via pandoc}}

\makeatletter
\newcommand\@shorttitle{}
\newcommand\shorttitle[1]{\renewcommand\@shorttitle{#1}}
\usepackage{fancyhdr}
\fancyheadoffset{0pt}
\makeatother
\renewenvironment{abstract}{
  \centerline
  {\large\sffamily\bfseries Abstract}\vspace{-0.25em}
  \begin{quote}\small
}{
  \end{quote}
}

\title{\sffamily\bfseries\huge\parfillskip=0pt
\rightskip=0pt plus .5\textwidth
\leftskip=0pt plus .5\textwidth
\emergencystretch=.3\textwidth The Role of Confounders and Linearity in
Ecological Inference: A Reassessment}
\shorttitle{Confounders and Linearity in Ecological Inference}
\author{\textbf{Shiro Kuriwaki}\footnote{
To whom correspondence should be addressed.
Email: \texttt{\href{mailto:shiro.kuriwaki@yale.edu}{shiro.kuriwaki@yale.edu}}.
We thank P. M. Aronow, Adam Chapnik, Gary King, Mason Reece, and Jesse
Shapiro for helpful comments.}
\\Department of Political Science%
\\Yale University%
\vspace{2pt}
 \and \textbf{Cory McCartan}
\\Department of Statistics%
\\Pennsylvania State University%
\vspace{2pt}
 }
\date{July 2026}

\begin{document}
\allsectionsfont{\sffamily}

\maketitle

\begin{abstract}
Estimating conditional means using only the marginal means available
from aggregate data is known as the ecological inference problem.
\begingroup\color{RevisionColor}We reassess this literature, arguing that it has understudied two issues: how practitioners should control for confounding, and how methodologists can leverage the linearity inherent in the structure of the problem.
On the former, we formalize ignorability conditions like those in causal inference and outline consistent plug-in estimators:
These are credible when covariates make the ignorability condition plausible.
On the latter, we show that aggregation restricts the target function to be partially linear.
Such linearity clarifies the connections between King's (1997) methodology, its predecessors, and subsequent developments.
That motivates a recent doubly-robust technique that enters covariates flexibly while leveraging linearity.\endgroup~Finally,
we test these methods in datasets where the ground truth is fortuitously
observed. In these common applications, all methods tested were prone to
overestimating racial polarization and underestimating split-ticket
voting.
\end{abstract}

\begin{refsection}

\section{Introduction}\label{introduction}

Estimating conditional means with only marginal means that come from
aggregate data is commonly known as the \emph{ecological inference}
problem (EI). These estimation challenges are central in many political
science applications, where for example voters' choices and
administrative data are aggregated to geographical districts.\footnote{Substantive
  questions that grapple with these ecological inference problems
  include: the degree of racially polarized voting
  \autocite{greiner2010exit}, the gender gap
  \autocite{teele2024political}, the prevalence of ticket splitting
  \autocite{burden2009americans}, the role of job loss on voting for
  Brexit \autocite{colantone2018global}, and partisan differences in
  pandemic exposure \autocite{NBERw34285}. While others use survey data
  to circumvent the ecological inference problem
  \autocite{baccini2021gone}, survey samples and self-reported vote
  choice may be unrepresentative. The same problem arises in economics
  (where consumers' purchasing choices for goods are aggregated into
  regional markets), public health (where residents' health outcomes are
  aggregated into census areas to preserve privacy), and other
  disciplines that use census statistics.} Researchers since
\textcite{robinson1950ecological} have been aware that such
relationships between aggregate data may not correspond to the
underlying relationship between individuals, calling incorrect
inferences an \emph{ecological fallacy}. They have produced a long
literature with statistical methods to make valid inferences from
aggregate data. Even after the publication of King's
\autocite*{king1997solution} \emph{A Solution to the Ecological
Inference Problem}, over twenty distinct sets of authors have proposed
methods and adjustments for ecological inference
(Appendix~\ref{sec-app-rev}).

Despite the range of existing methods, the concern persists that
aggregate data could result in an ecological fallacy. Some practitioners
take a cautious stance, refusing to make any inferences with aggregate
data. A wider community, including those beyond academia, uses the
existing EI methods regularly---EI is widely used to draw inferences
about racially polarized voting in Voting Rights Act cases in the
U.S.---and sometimes without interrogating the possibility of an
ecological fallacy. However, the conditions under which these fallacies
can occur are rarely articulated precisely.

In this paper, we provide a reassessment of ecological inference
methods.\footnote{ As we explain below in Section~\ref{sec-qoi}, we
  limit our review to EI methods that produce point estimates, as
  opposed to the partial-identification literature that focuses on
  bounding the unknown quantities of interest.} The existing
methodological literature on EI tends to treat EI as an almost unique
problem requiring unique solutions. We go beyond a mere review of this
past work, and instead provide a reformulation that explains the
ecological inference problem within a framework of
\begingroup\color{RevisionColor}missing data, causal inference, and linear regression modeling\endgroup.
By placing EI in this more general framework, users can rely on
intuition from statistical first principles on how and when ecological
fallacies occur. Our reassessment is also empirical. We evaluate both
well-established and new EI methods on common examples in political
science, and show why each tends to underestimate or overestimate the
quantity of interest.

Our paper consists of three major parts. After an illustrative example
of the ecological fallacy, we first define our quantities of interest
and the estimation challenge. The goal is to identify a conditional
expectation of an outcome \(Y\) conditional on a categorical predictor
variable \(X\), using data that has been coarsened into groups (often,
geographies) that contain a mix of categories of \(X\). For example, we
are interested in the population average of vote choice conditional on
racial group identification, but electoral districts coarsen observed
data into groups that are each a mixture of White voters, Black voters,
and Hispanic voters.

\begin{rrchange}

We show that a sufficient condition for identification is that the
expectation of the outcome, within individuals of a certain predictor
group, is independent of the prevalence of the predictor group, after
controlling for a set of observed covariates. Formally, this amounts to
a \emph{coarsening at random} or CAR condition
\autocite{heitjan1991ignorability}. Our particular use of CAR for
ecological inference is akin to selection on observables in causal
inference. Although many researchers have referred to a similar
identification condition over time
\autocite{king1997solution,hanushek1974model,glynn2008alleviating,chambers2001simple},
none to our knowledge have formalized the identification condition
nonparametrically and conditional on covariates, as we do here. All
methods of ecological inference fail to consistently estimate the
quantity of interest when this condition does not hold.

We further show in this reformulation that aggregation itself, inherent
in ecological data, provides additional valuable structure when
incorporating covariates: the outcome is always partially linear in the
categories of interest under CAR. This makes estimation for the
practitioner more straightforward: interact each covariate, possibly
after some transformation, with the predictor variable and run a linear
regression of the outcome on these interactions.

\end{rrchange}

In the second part of the paper, we re-characterize existing methods for
ecological inference in this framework. Although prominent existing
methods do not explain their models in this way, their models can be
represented by a certain linear regression as well. Our reformulation
shows that count models that were developed to handle more than two
outcome values, commonly known as \emph{R×C} EI methods, impose a
markedly different regression model than King's
\autocite*{king1997solution} original formulation, and introduce a new
set of conditions not previously understood. We also discuss how the
linear regression framework leads to novel methods that can not only
include confounders but are doubly robust to functional form
misspecification of the confounders \autocite{eimethods}.

In the third part of the paper, we explore how these methods perform in
practice. We use real datasets from two common use cases of ecological
inference in political science. We chose particular datasets that reveal
the ground truth quantity of interest, so that we can evaluate the
methods. In the example of estimating the partisan leanings of racial
groups, we show that ecological inference estimates tend to
\emph{overestimate} the degree of racial polarization. We show this is
partly because the types of Black voters who reside in neighborhoods
with higher proportions of Black residents disproportionately affect the
estimates for Black voters overall. In the example of estimating how
voters vote jointly across two offices, such as President and U.S.
House, we show that ecological inference tends to \emph{underestimate}
the prevalence of ticket splitting---voters who back different parties
for different offices. This is because ticket splitting voters are
scarcest in precisely the areas with high leverage in the regression.
These two cases feature different patterns, but the regression framework
we advance helps practitioners make sense of why these estimation errors
arise.

\section{An Example of the Fallacy}\label{an-example-of-the-fallacy}

An illustrative example of a classic ecological fallacy will help
foreground our formal treatment of the problem. In the U.S. presidential
election of 1968, George Wallace, the third-party candidate who had the
most segregationist policy platform in this election, won over a third
of the vote in the former Confederate states. Understanding the source
of voter support for Wallace by racial group is of interest to research
on American Politics, especially in the context of 1968, when the New
Deal coalition among White voters is thought to have fallen apart.

A single summary of the voteshare is reported in each geography, each of
which contains a mix of Black and non-Black voters. This makes it
impossible to back out the Wallace voteshare among Black and non-Black
voters separately in each geography. However, if the Wallace voteshare
is 1 point lower in a county that is 61\% Black vs.~an otherwise
equivalent county that is 60\% Black, we might infer that Black voters
do not vote for Wallace at all. A simple intuition for ecological
inference, then, is to plot the Wallace vote against the racial
composition, and extend the slope of the line to a hypothetical, 100\%
Black county.

\begin{figure}

\centering{

\includegraphics[width=0.7\linewidth,height=\textheight,keepaspectratio]{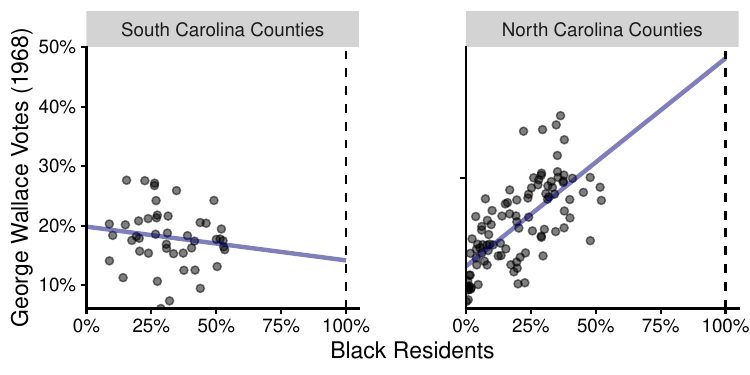}

}

\caption{\label{fig-wal}\textbf{Examples of the Ecological Fallacy}.
X-axis shows the Black population from the 1970 Census as a share of the
overall voting-age population (VAP). The y-axis shows the total number
of votes cast for George Wallace as a share of VAP. The estimated VAP
turnout in the two states was 54\%, which was in turn split across Nixon
(21\%), Wallace (17\%), and Humphrey (16\%).}

\end{figure}

Figure~\ref{fig-wal} shows such a scatterplot of these two observed
variables. In South Carolina, county-level voteshare for Wallace is
negatively correlated with the prevalence of Black voters.\footnote{This
  figure is complicated by differential turnout by race; see
  Appendix~\ref{sec-app-wallace-turnout}.} From this aggregate data one
might guess that individual Black voters do not support Wallace:
counties with fewer of these voters tend to have fewer Wallace votes.
However, in neighboring North Carolina, the relationship flips. Counties
with a higher Black population, covering the coastal Piedmont region,
are where Wallace wins the most votes. A naive observer might conclude
that Black voters were more likely to vote for Wallace than non-Black
voters in North Carolina.

Observers of these elections knew that such results could not be taken
to indicate Black voters preferred Wallace. Newly enfranchised Black
voters would have little reason to vote for Wallace, a candidate
explicitly running against the Civil Rights movement. Indeed, an
academic survey of this election showed that only 0.3\% of Black survey
respondents who reported voting in the former Confederate states
reported voting for Wallace \autocite{wright1977contextual,icpsr07508}.
Instead, they reasoned that White voters reacted to the Civil Rights Act
of 1964 and the Voting Rights Act of 1965, turned against the incumbent
Democratic administration and cast their votes for the Southern
candidate, Wallace
\autocite{wright1977contextual,phillips2014emerging,schoenberger1971ecology}.
The Wallace support among non-Black, White voters in North Carolina was
higher specifically in the areas where there were relatively more Black
voters, thus creating an ecological fallacy. The next section formalizes
this intuition.

\section{Confounding and Linearity in Ecological
Inference}\label{confounding-and-linearity-in-ecological-inference}

We clarify how certain variables confound the identification of the
estimand, and show how a linear model structure arises naturally from
the necessary identification assumption. This implies a certain linear
regression that accounts for potential confounders is the correct
approach to infer individual means from aggregate data. We conclude by
putting our identification result in the context of existing literature.

\begin{rrchange}

\subsection{Quantities of interest}\label{sec-qoi}

\end{rrchange}

The ecological inference problem is estimating the expectation of a
variable \(Y\) given a categorical variable \(X\) when individual
observations are averaged according to a grouping variable \(G\). We let
individuals belong to one of \(K\) categories of \(X\), and denote
\(X_{ik}\) to be a binary variable indicating whether individual \(i\)
belongs to category \(k\). We use the notation \(I\) to indicate a set
of individuals, so, for example, \(I_k\) indicates all individuals who
are of category \(k\). We use \(N\) to count these individuals, so
\(N_k\) is the number of individuals in the set \(I_k\). The outcome
\(Y\) is typically categorical in political science applications, and
represented as a vector of indicator variables, but it need not be.

The first quantity of interest is the conditional mean of the outcomes
among all individuals in category \(k\),
\begin{equation}\protect\phantomsection\label{eq-Bk}{
B_k := \frac{1}{N_k}\sum_{i\in I_k}Y_i.
}\end{equation} We may refer to this as the global parameter. Next, the
\emph{local parameters} are the conditional average for each geography
\(g\), \begin{equation}\protect\phantomsection\label{eq-bk}{
B_{gk} := \frac{1}{N_{gk}}\sum_{i\in I_{gk}} Y_i,
}\end{equation}

where \(I_{gk}\) again refers to individuals who are in both category
\(k\) and geography \(g\). To disambiguate between \(\vb X\) and \(G\),
we refer to levels of \(\vb X\) as categories (indexed by \(k\)) and
\(G\) as geographies (indexed by \(g\)), although in practice \(G\) can
represent non-geographic categories as well. The global parameter
\(B_k\) is exactly a weighted average of the local parameters \(B_{gk}\)
where the weights are the size of category \(k\) in each
geography.\footnote{ That is,
  \(B_k = (\sum_{g} N_{gk}B_{gk})/(\sum_{g} N_{gk})\) holds exactly. See
  the appendix for a derivation.}

A key identity relates these quantities of interest to the data and is a
core characteristic of the ecological inference problem. Using
\(\vb{B}_g\) to denote the \(K\)-length vector \([B_{g1}, ..., B_{gK}]\)
for each geography \(g\), we have

\begin{lemma}[Accounting
identity]\protect\hypertarget{lem-acct-id}{}\label{lem-acct-id}

For any geography \(g\), the aggregate outcome mean is exactly a linear
combination of the local conditional means:

\begin{equation}\protect\phantomsection\label{eq-acct-id}{
\bar{Y}_g = \vb{B}_g^\top \bar{\vb X}_g.
}\end{equation}

\end{lemma}

\begin{proof}
First notice that because the category \(\vb X\) is discrete, the sum of
outcomes \(Y\) in the entire geography can be partitioned into \(K\)
terms:
\(\sum_{i\in I_g} Y_i = \sum_{i\in I_{g1}} Y_i  + \cdots + \sum_{i\in I_{gK}} Y_i.\)
Then, each sum can be rewritten as a product of the composition
\(\bar{\vb X}\) and the local conditional means, \(\vb B_g\),
\begin{equation}\protect\phantomsection\label{eq-acct-id-dots}{
\begin{aligned}
\bar{Y}_g &= \frac{1}{N_g}\sum_{i\in I_g} Y_i\\
&= \frac{1}{N_g}\frac{N_{g1}}{N_{g1}}\sum_{i\in I_{g1}} Y_i + \cdots +
    \frac{1}{N_g}\frac{N_{gK}}{N_{gK}}\sum_{i\in I_{gK}} Y_i\\
&= \frac{N_{g1}}{N_g}B_{g1} + \cdots + \frac{N_{gK}}{N_g}B_{gK}\\
&= \bar{X}_{g1}B_{g1} + \cdots + \bar{X}_{gK}B_{gK}.
\end{aligned}
}\end{equation} The third line follows from Eq.~\ref{eq-bk}, and the
final line holds because the proportion of the geography \(g\) that is
also of category \(k\) is exactly the sample mean of \(X_{k}.\)
\end{proof}

This identity, which holds exactly in finite samples, is helpful for
subsequent estimation because it restricts the functional form of
possible models to one that is linear in the composition of categories.

\begin{rrchange}

Thus far, the global and local parameters are finite sample quantities.
Quantifying the statistical performance of our estimators requires that
we introduce a notion of randomness. Here, we conceptualize a
superpopulation from which geographies are drawn to define a
superpopulation version of our parameters, which we term the (global)
\emph{estimand}. The expectation of our conditional means over a
hypothetical infinite sample of geographies is
\begin{equation}\protect\phantomsection\label{eq-beta}{
\vbg\beta := \E[\vb B],
}\end{equation} where each element of \(\vbg\beta\) is
\(\beta_k = \E[B_k].\) This expectation, and all subsequent expectations
in this paper, is effectively an average over an infinite sample of
hypothetical geographies, and allows us to compute confidence intervals.
Practitioners may often be interested in \(\vb B\) itself, rather than
its superpopulation mean. When the number of geographies is large, the
difference between \(\vb B\) and \(\vbg\beta\) is negligible compared to
the variation in the \(\vb B_g\). Alternative conceptualizations of
randomness within the finite sample setting include a Bayesian approach,
which treats \(\vb B\) as a random variable with a prior rather than as
a parameter, and a design-based approach, which treats each individual's
choice of geography as the source of randomness.

\end{rrchange}

\begin{rrchange}

\subsection{Connection to missing data and causal
inference}\label{connection-to-missing-data-and-causal-inference}

\end{rrchange}

\begin{rrchange}

From this setup, it is clear that ecological inference is a missing data
problem: marginal means are observed while the conditional mean is not.
We propose that in particular, causal inference using potential outcomes
provides an intuitive and illuminating analogy for understanding the
subsequent identification conditions. Several aspects of the causal
setup map more cleanly to ecological inference than the missing data
setup does.

The key accounting identity in ecological inference is analogous to the
key relationship in causal inference. In causal inference with a binary
treatment \(D\) for unit \(i\), we have
\begin{equation}\protect\phantomsection\label{eq-causal-po}{
Y_i = Y_i(1)D_i + Y_i(0)(1-D_i),
}\end{equation} where the potential outcomes \(Y_i(0)\) and \(Y_i(1)\)
are unobserved and are effectively \emph{coarsened} into observed
\(Y_i\) by the treatment assignment \(D_i\). Causal estimands usually
involve expectations over the unobserved potential outcomes. For
example, the average treatment effect is
\(\E[Y(1) - Y(0)] = \E[Y(1)] - \E[Y(0)].\) The relationship in
Eq.~\ref{eq-causal-po} maps well to the accounting identity
(Eq.~\ref{eq-acct-id}) with two racial categories,
\begin{equation}\protect\phantomsection\label{eq-causal-ei}{
\begin{aligned}
\bar Y_g &= B_{g1} \bar X_{g1} + B_{g2} \bar X_{g2}\\
&=  B_{g1} \bar X_{g1} + B_{g2} (1 - \bar X_{g1}),
\end{aligned}
}\end{equation} where \(B_{g1}\) and \(B_{g2}\) are the unobserved local
parameters, which are coarsened into the local mean \(\bar Y_g\) by
aggregation based on the racial composition \(\bar X_{g1}\). These local
parameters can be viewed as the potential outcomes for geography \(g\)
if in fact its racial composition were homogeneously one group or the
other and voter behavior were held constant. And as in causal inference,
our quantities of interest, \(\beta_1 = \E[B_1]\) and
\(\beta_2 = \E[B_2]\), are averages of these unobserved potential
outcomes.

Thus, in both causal inference and ecological inference, we observe a
linear combination of unobserved or missing data, and hope to infer the
average value of the missing data across the population. The
\emph{fundamental problem of causal inference} embodied in
Eq.~\ref{eq-causal-po} is that there are twice as many unobserved
potential outcomes as there are individuals, so the estimand is
unidentified without further assumptions. Eq.~\ref{eq-causal-ei} makes
clear that ecological inference also involves attempting to identify two
parameters for every data point.

The main difference between the two problems is also illuminating: in
the causal setup above, treatment \(D_i\) is binary, whereas in
ecological inference, the corresponding variable \(\bar X_{g1}\) is a
proportion. This is not to say that EI is causal inference with a
continuous treatment: with binary \(D_i\) exactly one potential outcome
is observed for each individual. A continuous treatment would require a
different potential outcome for each of the infinitely-many treatment
values. In EI, we observe a continuous mixture of a finite number of
unobserved \(B\) for each geography. The continuous nature of this
mixture pays dividends in carrying out estimation, and in understanding
threats to inference, as we discuss below.

\end{rrchange}

There are other related ways in which the causal inference analysis is
fruitful. The conditions needed to identify the ecological inference
estimands are similar to the familiar ignorability conditions in causal
inference. As we show below, we can therefore motivate the importance of
controlling for confounders and positivity conditions in the same way as
in causal inference designs.

\subsection{Identifying the global
estimand}\label{identifying-the-global-estimand}

The presence of more unobserved missing values than observations means
that neither causal nor ecological estimands can be identified without
further assumptions. In causal inference, this problem is often tackled
by making an \emph{ignorability} assumption that the missing potential
outcomes are unrelated to the treatment assignment. This assumption is
satisfied in a randomized experiment.\footnote{Assuming full treatment
  compliance, no spillover, and so on.}

The analogous assumption in ecological inference is also sufficient for
identification, with a small twist: we must also consider how the
unobserved \(\vb B_g\) relate to \(N_g\), the number of individuals in
each geography. Specifically, if
\begin{equation}\protect\phantomsection\label{eq-ccar}{
\E[\vb B_g | \bar{\vb X}_g=\bar{\vb x}, N_g = n] = \E[\vb B]\quad\text{for every value of}\ \bar{\vb x}, n,
}\end{equation} we say that the assumption of \emph{coarsening
completely at random} (CCAR) holds. The following proposition shows that
this assumption is sufficient to identify \(\vbg\beta =  \E[\vb B]\).
The proof is in Appendix~\ref{sec-app-proofs}.

\begin{proposition}[Identification under
CCAR]\protect\hypertarget{prp-id-ccar}{}\label{prp-id-ccar}

If CCAR holds, then \(\vbg\beta\) is identified as the (population)
regression coefficients of \(\bar Y\) on \(\bar{\vb X}\), \[
\vbg\beta = \E[\bar{\vb X}\bar{\vb X}^\top]^{-1} \E[\bar{\vb X}\bar Y].
\]

\end{proposition}

In other words, a regression of \(\bar Y_g\) on \(\bar{\vb X}_g\) with
no intercept can consistently estimate \(\vbg\beta\), as long as one
believes that CCAR holds. In the ecological inference context, this
estimator is known as \emph{Goodman regression} or \emph{ecological
regression}.\footnote{\textcite{goodman1953ecological} itself, however,
  warned against making a CCAR-like constancy assumption
  \autocite{wakefield2004ecological}.}

\begin{rrchange}

Compare CCAR to the mean-independence condition (or weak ignorability)
used to estimate the average treatment effect in causal inference. That
condition states that the treatment variable is mean-independent with
the missing data, i.e., \(\E[Y(d) | D_i = d] = \E[Y(d)]\) for \(d = 0\)
and \(d = 1\). This matches the CCAR assumption, where the assignment of
voters to locations, which generates \(\bar{\vb X}\) and \(N\), plays
the role of the treatment assignment \(D\). If weak ignorability holds,
then a simple regression of \(Y\) on \(D\)---which is mathematically
identical to a difference in means between the treated and control
groups---consistently estimates the average treatment effect. Thus,
Goodman's regression can be thought of as the analog to the causal
difference-in-means estimator.

\end{rrchange}

Is coarsening completely at random plausible in practice? Consider again
the 1968 election example. In this case, CCAR means that the average
Wallace support among White voters in heavily White areas is equivalent
to that in other areas, or the population as a whole. For instance, this
means White voters in rural lowland counties with a higher proportion of
Black voters and White voters in mountainous counties with few racial
minorities support Wallace at the same level in expectation. CCAR is
implausible in this setting, and is incompatible with the social
scientist's interest in how different individuals sort into different
geographies. This helps explain why the simple linear regression shown
in Figure~\ref{fig-wal} for North Carolina has the wrong slope.

If CCAR is implausible, what is to be done? In causal inference, an
assumption of complete randomization can be weakened to hold conditional
on covariates, which is known as a \emph{selection-on-observables}
assumption. The same idea applies to ecological inference: covariates
can solve the ecological fallacy under certain conditions.

Suppose there exists a variable \(\vb Z_g\) (in general, a vector) that
is observed at the geography level. For example, \(Z\) can be the income
in a geography, or a binary variable indicating whether the geography is
a certain region of the state. Then, if
\begin{equation}\protect\phantomsection\label{eq-car}{
\E[\vb B_g\mid \vb Z_g = \vb z, \bar{\vb X}_g = \bar{\vb x}, N_g = n] = \E[\vb B_g\mid \vb Z_g = \vb z]\quad\text{for every value of}\ \vb z, \bar{\vb x}, n,
}\end{equation} we say that \emph{coarsening at random} (CAR)
holds.\footnote{ Note that compared to \textcite{imai2008bayesian}, who
  discuss similar assumptions, our acronyms are reversed: there, CAR is
  the stronger assumption (our CCAR), and CCAR (where the first C stands
  for ``conditional'') is the weaker assumption. We have opted for
  CCAR/CAR here to match the existing MCAR/MAR terminology in the
  missing data literature.} The intuition of this equation is that among
geographies with a particular set of covariate features
\(\vb Z_g=\vb z\), knowing \(\bar{\vb X}_g\) and \(N_g\) does not change
the expected value of \(\vb B_g\). CCAR is a special case of CAR where
\(\vb Z_g\) contains no covariates at all. Versions of this assumption
have been discussed, often informally, in the literature, a history we
review in Section~\ref{sec-id-hist} below. This assumption is sufficient
to identify the quantity of interest \(\vbg\beta\), as the following
proposition formalizes.

\begin{proposition}[Identification under
CAR]\protect\hypertarget{prp-id}{}\label{prp-id}

For all categories \(k\), if coarsening at random holds, \(\vbg\beta\)
is identified as\footnote{
\begingroup\color{RevisionColor}
  Following convention and the previous proposition, we suppress the subscript $g$ for variables inside superpopulation expectations. 
  But we make an exception for the denominator $\E[N_{gk}]$ in order to distinguish it from the total number of category-$k$ individuals, $N_k$.
  The ratio $N_{gk} / \E[N_{gk}]$ represents how much geography $g$'s category-$k$ population differs from the typical geography, and merely re-weights the geographies to represent individuals instead of geographies.
  When every geography contains the same number of category-$k$ individuals, the proposition is that $\beta_k = \E[\E[\bar Y\mid \vb Z = \vb z, \bar X_{k}=1]].$
\endgroup
} \[
\beta_k = \E\left[\E[\bar Y\mid \vb Z = \vb z, \bar X_{k}=1] \frac{N_{gk}}{\E[N_{gk}]}\right].
\]

\end{proposition}

At first glance, Proposition~\ref{prp-id} may appear very different from
Proposition~\ref{prp-id-ccar}. However, when CCAR holds and \(\vb Z\) is
empty,
\(\E[\bar Y\mid \vb Z, \bar X_{k}=1]=\E[\bar Y\mid \bar X_{k}=1]\) is
constant, and so the identification expression simplifies to
\(\beta_k=\E[\bar Y\mid \bar X_{k}=1]\). The value of the Goodman
regression when \(\bar X_k=1\) and all other \(\bar X_{k'}=0\) is
exactly the coefficient on \(\bar X_k\), which is the identification
result in Proposition~\ref{prp-id-ccar}.

Just as Proposition~\ref{prp-id-ccar} implies a certain natural plug-in
estimator (Goodman's regression), Proposition~\ref{prp-id} does as well.
To estimate the conditional average for category \(k\):

\noindent \textbf{Procedure 1}

\begin{enumerate}
\def\labelenumi{\arabic{enumi}.}
\tightlist
\item
  Fit a regression model of \(\bar Y\) on some function of
  \(\bar{\vb X}\) and \(\vb Z\).
\item
  For each observation in the regression, produce fitted values
  \(\hat{y}\) on a hypothetical dataset where all geographies contain
  only individuals of category \(k\), such that \(\bar X_{gk}=1\) and
  \(\bar X_{gk'}=0\) for the other categories \(k'\neq k,\) while other
  values \(\vb Z\) are held at their observed values,
\item
  Then \(\hat{\beta_k}\) is the average over these fitted values
  \(\hat{y}\), weighted by \(N_{gk}\), the number of individuals in
  category \(k\) in each geography \(g\).
\end{enumerate}

The plug-in and aggregation steps are straightforward, but what is
challenging is the first step of estimating the unknown CEF
\(\E[\bar Y\mid \vb Z = \vb z, \bar{\vb X}_g = \bar{\vb x}].\) We
discuss the aspects of this problem in several steps, starting with a
simple functional form (Section~\ref{sec-linear-inter}), increasing
robustness to functional form misspecification
(Section~\ref{sec-semipara}), and accounting for unobservable
confounders (Section~\ref{sec-empirics}).

\subsection{Identifying the local
estimand}\label{identifying-the-local-estimand}

The accounting identity also clarifies that, unfortunately, point
identification of the local \(\vb B_g\) is impossible without further
assumptions: each observation contributes \(K\) unknown parameters (the
entries of \(\vb B_g\)), so there are \(K\) times as many parameters as
observations. Existing literature has taken one of two approaches to
this challenge.

\begin{rrchange}

The accounting identity allows for \emph{partial identification}, that
is, estimating strict bounds on the possible values the unobserved
parameters can take. As an extreme example, if geography \(g\) is
composed exclusively of category \(k\), the local estimand \(B_{gk}\)
must be exactly the observed outcome \(\bar{Y}_g\). In general, when
\(Y\) is a bounded variable, the quantities of interest \(B_{gk}\) are
bounded by a function of the compositions and the outcome, a constraint
first noted by \textcite{duncan1953alternative}.
\textcite{king1997solution} combined these Duncan--Davis bounds with the
Goodman regression framework, and others have explored how these bounds
generalize for other quantities of interest
\autocite{cross2002regressions} or how they can be narrowed with
additional assumptions \autocite{jiang2020ecological,NBERw34285}.

\end{rrchange}

Our paper does not focus on partial identification so that we keep the
focus on the bias in point estimates. However, the Duncan--Davis bounds
contain information that is certainly helpful in their own right. In
some extensions, the bounds can be sufficiently narrow to be useful, and
bounds may also test the plausibility of the identifying assumptions we
introduced above: if, even with many observations, the point estimates
are outside the bounds, then the identifying assumptions are likely
violated.

\begin{rrchange}

Another approach is to take a Bayesian perspective and treat the
\(\vb B_g\) as parameters. While a consistent estimate of each
\(\vb B_g\) is still impossible, the Bayesian approach at least allows
uncertainty quantification for these local estimands. This is the
approach that \textcite{king1997solution} and follow-up work take, as we
elaborate in more detail below. In particular, King constructed the
prior on \(\vb B_g\) hierarchically and in a way that incorporated the
bounds on \(Y_g\), so that the posterior estimates of \(\vb B_g\) lie
within the Duncan--Davis bound automatically. A challenge with this
approach is the sensitivity to the particular prior distribution chosen,
and especially the assumptions that distribution makes about
correlations between the elements of \(\vb B_g\).\footnote{ In the
  framework of \textcite{king1997solution}, the covariance of the
  entries in \(\vb B_g\) is what specifies the angle at which the global
  estimate is projected down to the tomography line to make a point
  estimate.} This is the same challenge that arises in Bayesian causal
inference, where sample-level causal estimands depend on untestable
assumptions about the joint distribution of the potential outcomes.

However, unlike causal inference, it may be possible to make some
inferences about the distribution of \(\vb B_g\) from the data.
\textcite{eimethods} show that the covariance of the local estimands can
be identified under a second-moment independence condition conditional
on covariates, i.e., that
\(\E[\vb B_g \vb B_g^\top \mid \vb Z_g, \bar{\vb X}_g, N_g] = \E[\vb B_g \vb B_g^\top \mid \vb Z_g].\)
This is a similar but stronger assumption to CAR. They use this
additional assumption to construct asymptotically valid confidence
intervals for the local estimands. While these intervals will not shrink
to zero as more data are collected, due to the fundamental
unidentifiability, they provide useful uncertainty quantification, just
as Bayesian approaches to ecological inference aim to do.

\end{rrchange}

\subsection{The role of linearity}\label{sec-linear-inter}

Proposition~\ref{prp-id} shows that enough covariates can identify
ecological inference parameters, yet the question of how to estimate
these quantities remains, because the functional form with which
collected covariates interact is unknown. Fortunately, the accounting
identity Eq.~\ref{eq-acct-id}, combined with CAR, restricts the true
regression function of \(\bar Y\) on \(\vb Z\) and \(\bar{\vb X}\) to be
\emph{partially linear} in \(\bar{\vb X}\). This proves helpful in
estimation.

Why does partial linearity arise in ecological inference specifically?
Note that the CAR assumption implies the following structure for
\(B_g\): \begin{equation}\protect\phantomsection\label{eq-bg-decomp}{
\begin{aligned}
\vb B_g &= \E[\vb B_g\mid \vb Z_g, \bar{\vb X}_g] + \vbg\eps_g\\
&= \E[\vb B_g\mid \vb Z_g] + \vbg\eps_g \quad\text{with}\quad \E[\vbg\eps_g\mid \bar{\vb X}_g] = 0
\end{aligned}
}\end{equation} where the second line is directly due to CAR in
Eq.~\ref{eq-car}. The residuals \(\vbg\eps_g\) are conditionally
mean-zero because the residuals from any conditional expectation are
orthogonal to the conditioning variables. Eq.~\ref{eq-bg-decomp} merely
re-expresses the CAR assumption in a form that will prove more
convenient, but it highlights a crucial implication of CAR: on average,
the local estimand is a function only of \(\vb Z_g\), with residual
variance unrelated to \(\bar{\vb X}_g\).

Now, substituting Eq.~\ref{eq-bg-decomp} into Eq.~\ref{eq-acct-id} and
denoting \(\E[\vb B_g\mid \vb Z_g]\) as \(f(\vb Z_g),\) some (unknown)
vector-valued function of the covariates, we have that \[
\bar Y_g = (f(\vb Z_g) + \vbg\eps_g)^\top \bar{\vb X}_g
= f(\vb Z_g)^\top \bar{\vb X}_g + \vbg\eps_g^\top \bar{\vb X}_g.
\] Taking conditional expectations of both sides, we find
\begin{equation}\protect\phantomsection\label{eq-cef}{
\begin{aligned}
\E[\bar Y_g \mid \vb Z_g, \bar{\vb X}_g]
&= f(\vb Z_g)^\top \bar{\vb X}_g \\
&= f_1(\vb Z_g)\bar{X}_{g1} + \cdots + f_K(\vb Z_g)\bar{X}_{gK},
\end{aligned}
}\end{equation} because the residual term \(\vbg\eps_g\) is
conditionally mean-zero. The left-hand side of Eq.~\ref{eq-cef} is the
conditional expectation function (CEF) of \(\bar Y_g\) on \(\vb Z_g\)
and \(\bar{\vb X}_g\), which appears in Proposition~\ref{prp-id}. What
the right-hand side of Eq.~\ref{eq-cef} shows is that the CEF has a
partially linear structure: it is linear in \(\bar{\vb X}_g\) with
coefficients that depend, possibly nonlinearly, on the covariates
\(\vb Z_g\). This type of model is also known as a varying coefficient
model \autocite{hastie1993varying,fan1999varying}.

Now, to make the exposition in this section more concrete, suppose we
are further willing to assume the coefficient functions \(f(\vb Z_g)\)
(that is, \(\E[\vb B_g\mid\vb Z_g]\)) are simply linear in \(\vb Z_g\).
Let \(p\) be the number of covariates in \(\vb Z_g\), then we can write
such a model as \[
f_k(\vb Z_g) = \gamma_{k0} + \gamma_{k1}Z_{g1} + \cdots + \gamma_{kp}Z_{gp},
\] where the regression coefficient \(\gamma_{kj}\) is indexed by each
predictor category \(k\) and the covariate index
\(j \in \{0, 1, \ldots, p\}.\)

\begin{rrchange}

We refer to this assumption as a \emph{linearity-in-covariates}
assumption (not to be confused with the CEF being always linear in
\(\bar{\vb X}\)). This is a strong assumption---in reality, each
covariate could enter \(f\) in a nonlinear form, or interact with other
covariates---but it helps illuminate what linearity-in-\(\bar{\vb X}\)
buys us. We relax this assumption in Section~\ref{sec-semipara}.

\end{rrchange}

Substituting this expression into Eq.~\ref{eq-cef}, we have:
\begin{equation}\protect\phantomsection\label{eq-cef-lin}{
\begin{aligned}
\E[\bar Y_g \mid \vb Z_g, \bar{\vb X}_g]
&= \underbrace{\gamma_{10}\bar{X}_{g1} + \gamma_{11}Z_{g1}\bar{X}_{g1} + \cdots + \gamma_{1p}Z_{gp}\bar{X}_{g1}}_{\text{For outcome category 1}} + \cdots \\
&\quad+~ \underbrace{\gamma_{K0}\bar{X}_{gK} + \gamma_{K1}Z_{g1}\bar{X}_{gK} + \cdots + \gamma_{Kp}Z_{gp}\bar{X}_{gK}}_{\text{For outcome category}~K}.
\end{aligned}
}\end{equation}

\begin{rrchange}

This CEF is fully linear, and so in this case an ordinary least squares
regression is an appropriate estimator for these coefficients. In other
words, to estimate \(\vbg\beta\), regress the outcome on pairwise
interactions of each of the \(p\) covariates \(\vb Z_g\) with the \(K\)
categories \(\bar{\vb X}\), then generate estimates following Procedure
1. Usually, the modeler faces the problem of not knowing how the
variables \(\bar{\vb X}\) and \(\vb Z\) should enter the regression:
whether they should be logged, squared, binned, or interacted. In
ecological inference problems, there is no such ambiguity for
\(\bar{\vb X}\): it should be fully interacted with \(\vb Z\) and enter
the regression linearly.

We consider Eq.~\ref{eq-cef-lin} to be a generalization of Goodman's
regression to the case where covariates are included. \footnote{That is,
  if \(\vb Z_g\) is empty because CCAR holds, then
  \(f(\vb Z_g)=\vbg\beta\) and Eq.~\ref{eq-cef} exactly expresses
  Goodman's regression.} Though Eq.~\ref{eq-cef-lin} requires a
linear-in-covariates assumption about \(f(\vb Z_g),\) even without it,
Eq.~\ref{eq-cef} follows immediately from the CAR assumption alone.

\end{rrchange}

\subsection{Implications of linearity}\label{implications-of-linearity}

The connection between aggregation and linear regression under CAR is
useful beyond justifying the use of linear regression. The connection
allows us to use familiar intuition about linear regression to
understand tradeoffs in ecological modeling and anticipate when
ecological inference will fail.

\begin{figure}

\centering{

\includegraphics[width=0.8\linewidth,height=\textheight,keepaspectratio]{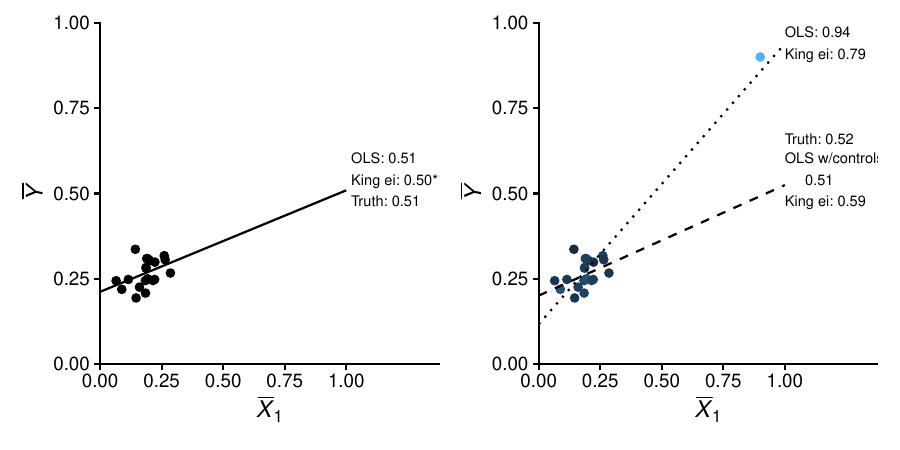}

}

\caption{\label{fig-sim}\textbf{Intuition for EI as linear regression.}
A simulated example where the quantity of interest is \(\beta_1=0.5\).
In panel \textbf{(a)}, \(n = 20\) data points are simulated from the
model of \textcite{king1997solution}. A simple OLS fit evaluated at
\(\bar X_1=1\) provides an estimate that agrees with both King's EI
algorithm and the ground truth. In panel \textbf{(b)}, an outlier with
high leverage is added to the dataset, shown in the top-right part of
the figure. The resulting OLS fit severely overestimates the truth, as
does, to a lesser extent, King's EI. However, if the outlier differs
with the other \(n=20\) points on some covariate \(Z\), controlling for
\(Z\) in the OLS reduces the bias significantly.}

\end{figure}

To demonstrate these connections, we introduce in Figure~\ref{fig-sim}
(a) a simple simulated example. We generated \(20\) synthetic precincts
with two racial groups that satisfy the CCAR assumption, with the global
parameter \(\vbg\beta\) set to \((0.5, 0.2)\).\footnote{ Each precinct's
  \(\vb B_g\) were drawn from a tightly distributed bivariate truncated
  normal distribution as in the model of \textcite{king1997solution}.}
In this sample, \(B_{1}\) was \(0.508\). We then fit an OLS regression
of the implied \(\bar{Y}\) on \(\bar{\vb X}\). The line evaluated at
\(\bar{X}_1\) in this particular example was \(0.509\), quite close to
\(\beta_1\) and even closer to the finite-sample \(B_1\). We also ran
King's \autocite*{king1997solution} model on the same \(20\) data
points, as a preliminary illustration of our point that King's estimator
can be represented by a particular Goodman regression. Its estimate for
\(B_1\) was \(0.511\), also quite similar. This example illustrates
Proposition~\ref{prp-id-ccar} that, under CAR (or CCAR), estimates of
\(\vbg\beta\) can be obtained by plugging in \(\bar{X}_1=1\) and
\(\bar{X}_1=0\) into a simple regression.

The near-perfect accuracy of the ecological inference in
Figure~\ref{fig-sim} (a) is, however, fragile. There are at least three
major ways in which we can characterize the direction of potential
errors.

\subparagraph{Fragility in
extrapolation}\label{fragility-in-extrapolation}

In many EI examples involving geographic units, perfect residential
sorting (\(\bar{X}_{k} = 1\) or \(0\)) is rare, so producing the plug-in
estimates from Procedure 1 usually involves inevitable extrapolation
from the observed data. This is clearly seen in Figure~\ref{fig-sim}
(a), where the observed \(\bar{X}_1\) are clustered near zero. As a
result, the Goodman estimate for this category is unstable, because much
extrapolation is required. Extrapolations can lead to even impossible,
out-of-bounds estimates that are negative or above 1. Finite-sample
variation combined with extrapolation can produce negative estimates
\emph{even when CCAR holds}. This feature of the estimator explains why
EI applications typically find that estimates for small minority groups
are highly variable or impossible.

\subparagraph{Influence and high-leverage
points}\label{influence-and-high-leverage-points}

Intuition about high-leverage or influential points in linear regression
also carries over to ecological inference.\footnote{ An influential
  point is an observation that has a disproportionate effect on the
  slope of a regression
  \autocite{blackwell2025user,chatterjee1986influential}. Specifically,
  influence is computed by the leverage of a point multiplied by its
  outlier value from the leave-one-out regression.} In
Figure~\ref{fig-sim} (b) we illustrate the influence of such a point by
adding a single, high-influence observation to the sample. This precinct
has \(\bar{X}_1 = 0.9\). It also has a different outcome than the other
distribution: \(B_{g1} = 1\), \(B_{g2} = 0,\) resulting in
\(\bar{Y}_g = 0.9\). The Cook's distance of this observation (a standard
measure of influence) is \(34.6\), compared to less than \(0.2\) for all
other observations. As a result, the slope of the OLS line changes
drastically, and the extrapolated estimate at \(\bar{X} = 1\) is now off
by over 40 percentage points, even though the global parameter \(B_1\)
only increased to \(0.53\). \textcite{king1997solution}'s estimate of
the same data is less drastically off, but not much better. Both methods
fail on panel (b) because of the violation of CCAR: the geography with
an especially large \(\bar{X}\) is also the one with an especially high
\(B_1\) value. The lesson here is that the connection to regression
serves as a reliable diagnostic to anticipate or identify possible
violations.

Fortunately, the use of control variables can reduce the undue impact of
influence points. In the same example in Figure~\ref{fig-sim} (b), if
there is a third variable \(Z\) that distinguishes this influence point
from the remaining \(20\) observations, the regression may be
corrected.\footnote{ In this example, we suppose that \(Z = 1\) for the
  influence point and it is a small random Normal error centered at
  \(0\) and with a standard deviation of \(0.001\) for the other \(20\)
  observations.} Applying the Procedure 1 to the data assuming linearity
in the single covariate generates a linear fit that is \(0.51\) when
evaluated at \(\bar{X}_1=1\) and \(Z=0\), as shown by the dashed line in
the figure. The covariate helps satisfy the CAR assumption while
decreasing the residual variance of the regression, and the overall
estimate is much improved.

\subparagraph{Positivity violations}\label{positivity-violations}

Finally, the linear regression intuition is also helpful in highlighting
the need for \emph{positivity}: sufficient variation in \(\bar{\vb X}\),
including residual variation after accounting for covariates. This is
the EI analogue of the overlap assumption in causal inference, which
requires that each individual has a nonzero probability of receiving
each type of treatment.

The precision in a linear regression estimate is generally increasing in
the variation in the predictor variable. Small variation in
\(\bar{\vb X}\) thus poses an estimation problem even if CCAR holds. For
example, estimates of the gender gap in elections are typically unstable
because gender ratios in most geographies are near parity, which means
little variation (not to speak of the significant extrapolation to a
hypothetical single-sex geography) \autocite{teele2024political}. In
regressions with covariates, the relevant identifying variation is the
residual variation in \(X\) remaining after holding covariates
fixed.\footnote{ Precisely, the coefficient on one variable (\(X\)) can
  be obtained by a sub-regression of the outcome residualized by the
  other covariates (\(Z\)) on that one variable (\(X\)) residualized in
  the same way, by the Frisch-Waugh-Lovell theorem.} In the extreme case
where \(\vb Z\) is collinear with \(\bar{\vb X}\) and thus explains all
of its variance, then the regression model cannot be estimated even
though CAR technically holds.\footnote{ Notice that CAR trivially holds
  if we let \(\vb Z=\bar{\vb X}\). But this trick cannot estimate
  \(\vbg\beta\) due to its complete collinearity.}

This poses a tradeoff: including control variables will improve the
plausibility of CAR but may also remove variation in \(\bar{\vb X}\)
necessary for estimation. For example, return to the 1968 election in
Figure~\ref{fig-wal} and imagine controlling for the proportion of the
population in each county that was enslaved in 1860. This variable is a
strong predictor of the proportion of Black voters in a county and also
of the voteshare for Gov.~Wallace. It is more plausible to believe that,
among counties with a similar history of slavery, White voters' support
for Wallace is unrelated to the proportion of Black voters. However, the
near-collinearity of the covariate with \(\bar X_{g1}\) means that the
regression will be highly unstable.

Fundamentally, ecological inference requires exogenous variation in
\(\bar {\vb X}\), from which estimates of \(Y\) for each category can be
extrapolated. When the covariates needed to satisfy CAR also mop up all
of the variation in \(\bar {\vb X}\), then there is simply not enough
information left in the data to make an estimate.

\subsection{On the history of controlling for
confounders}\label{sec-id-hist}

\begin{rrchange}

Many methodologists have been aware of the sort of conditional
identification condition we show here. \textcite{goodman1959some}
himself stated the CCAR assumption and considered several ways
covariates could make CAR hold {[}612, 622--644{]}.
\textcite{hanushek1974model} proposed a linear regression with
covariates, but included no interactions.

\textcite{king1997solution} presents a conditional independence
assumption briefly in its chapter on linear contextual effects and
avoiding aggregation bias, but does not formalize an identification
result {[}170--171{]}. \textcite{imai2008bayesian} identifies the need
for a coarsening at random assumption conditional on covariates, but
relies on a particular distributional model for the
parameters.\footnote{Neither of these sets of authors considers the role
  of \(N_g\) in the identification assumption.} Other work has focused
on model misspecification and bias under no covariates
\autocites[e.g.][]{ansolabehere1995bias,tam2004limits} while rarely
discussing the use of covariates. This direction in the literature is
notable since, as \textcite[175]{lewis2001understanding} observes,
``many scholars have considered aggregation bias (the violation of
{[}the constancy{]} assumption) to be \emph{the} problem in making
ecological inferences'' (emphasis ours).

\end{rrchange}

Controlling for covariates may have sat uneasily with the desire to not
control out intermediate mechanisms. This logic may have been reinforced
by court rulings in the late 80s that focused the research question on
the mere existence of racially polarized voting, regardless of its
underlying causes.\footnote{\emph{Thornburg v. Gingles}, 478 U.S. 30
  (1986).} Some scholars took this to mean that one should not control
for covariates in ecological inference \autocite[171]{king1997solution}.
The use of covariates also went against \textcite{achen1995cross}, which
argued that the biases of ecological regression, ``are not curable by
\ldots{} controlling for demographic variables'' at the individual level
(92--94). Our theoretical results clarify that one should control for
covariates such that CAR holds (a statement about aggregate level
ignorability, not individual level predictive accuracy), and then
combine the estimated \emph{ceteris paribus} coefficients into the
conditional mean of interest.

\section{Comparison of Methods for Ecological Inference}\label{sec-est}

\begin{rrchange}

Our regression framework lets us re-assess the differences across the EI
methods commonly used by practitioners. Existing models achieve
identification using one of two approaches: A \emph{parametric}
approach, which relies on specific distributional or functional form
assumptions controlled by a finite number of parameters, or a
\emph{non-parametric} approach, which does not rely on such assumptions
for accurate inference.

\end{rrchange}

\subsection{King's 2×2 ecological inference
model}\label{kings-22-ecological-inference-model}

One achievement of King's 1997 model is that it provides point estimates
of the local means \(\vb B_g.\) Starting from the accounting identity
with \(K=2\) categories and a binary \(Y\), King moves to a parametric
Bayesian regression in which the coefficients \((B_{g1}, B_{g2})\) are
jointly drawn from a bivariate Normal distribution truncated to the unit
square: \[
\vb B_g \iid \Norm_{[0,1]^2}(\vbg\mu, \Sigma),
\] where \([0, 1]^2\) denotes truncation to a unit square, \(\vbg \mu\)
is the location parameter, and \(\Sigma\) is the covariance matrix. The
truncation is applied so that the \(\vb B_g\) remain properly between
\(0\) and \(1\), since King's model applies only to binary \(Y\).

The estimation of each local estimand is best understood by first
considering the easier setting of when \(\vb B_g\) is untruncated. Such
a model is well-studied as a random coefficient model
\autocites{griffiths1972estimation}[as cited in][]{anselin2002spatial}.
In this case, the MLE for the local estimand for each geography \(g\),
conditional on \(\Sigma\), is given by
\begin{equation}\protect\phantomsection\label{eq-king-untrunc}{
\hat{\vb B}_g = \underbrace{\hat{\vb B}}_{\text{global estimate}} +\ \Sigma\bar{\vb X}_g(\bar{\vb X}_g^\top\Sigma\bar{\vb X}_g)^{-1}\underbrace{(\bar{Y}_g - \bar{\vb X}_g^\top\hat{\vb B})}_{\text{residuals}}.
}\end{equation} That is, the estimates are the combination of the global
estimate and the residual of the linear regression reweighted by the
covariances of the two coefficients. Intuitively, King's model produces
unit-level estimates by allocating the residuals from a certain weighted
Goodman regression in accordance with an estimate of the covariance
matrix \(\Sigma\).

The implications for identification remain the same in Goodman and
King's models. We can rewrite King's model for \(\vb B_g\) in terms of
the global parameter \(\vb B=\E[\vb B_g]\) and a residual term
\(\vbg\eps_g\). To do so, define \(\vb m_g = \vbg\mu - \vb B\). Unlike
with an untruncated Normal, \(\vb m_g\) is not always \(0\), because the
truncation can shift the mean of the distribution away from its location
parameter \(\vbg\mu\). Then we can write
\begin{equation}\protect\phantomsection\label{eq-bg-m}{
\vb B_g = \vb B + \vbg\eps_g, \quad
\vbg\eps_g \iid \Norm_{[0,1]^2 - \vb B}(\vb m_g, \Sigma),
}\end{equation} where \([0,1]^2 - \vb B\) is the unit square shifted by
the vector \(\vb B\). Since \(\vbg\eps_g\) is drawn independent of
\(\bar{\vb X}\), by construction, we have
\(\E[\vbg\eps_g \mid\bar{\vb X}_g]=0\), so
\(\E[\vb B_g \mid\bar{\vb X}_g]=\vb B\). This is exactly the CCAR
assumption. Thus \textcite{king1997solution}'s model makes the same
strong assumption as Goodman's regression for identification. Since the
model does not involve covariate adjustment, it does nothing to
ameliorate aggregation bias due to confounders
\autocite{rivers1998solution,lewis2001understanding}.\footnote{\textcite{king1997solution}
  defines aggregation bias in a looser way. Chapter 9 of the book
  discusses covariate adjustment, but also appears to consider
  out-of-bounds estimates as a type of aggregation bias. It is true that
  the truncation imposed here ultimately reduces estimation error in
  many cases. However, we agree with Lewis'
  \autocite*{lewis2001understanding} (175) interpretation that King's
  main method ``should be thought of as a `solution' in the sense that,
  assuming its assumptions hold, it allows the user to make more
  efficient estimates {[}of the global parameter{]} than can be made
  using conventional regression techniques and also allows the
  estimation of {[}local parameters{]}.''}

Substituting the reexpression into each geography's accounting identity,
we have \begin{equation}\protect\phantomsection\label{eq-regr-king}{
\bar Y_g = \vb B^\top \bar{\vb X}_g + \bar\eps_g,
}\end{equation} where
\(\bar\eps_g = \vbg\eps_g^\top \bar{\vb X}_g.\)\footnote{We replace the
  local coefficients in the accounting identity with Eq.~\ref{eq-bg-m},
  and rearrange in terms of the global parameter.} This is exactly the
form of the CEF that follows from the CCAR assumption (see also
Eq.~\ref{eq-bg-decomp}), on which Goodman's regression is based. The
difference is that Goodman's regression makes only the assumption that
\(\E[\bar\eps_g \mid \bar{\vb X}_g]=0\), whereas King's model assumes a
specific distribution for the error term \(\bar\eps_g\). This
distribution is not the same for every observation, and in fact depends
on the coefficients \(\vb B_g\), the same way that the error term in a
generalized linear model depends on the linear predictor. Still, this
added assumption often leads to improved estimation in finite samples
relative to untruncated regression because information on each
accounting identity is incorporated for each local estimate, and the
global estimates are produced as a mean of these local estimates.

This model, however, requires more advanced computational techniques
than linear regression: because \(\bar{\eps}_g\) is a linear combination
of (correlated) truncated Normal variables, it is not even truncated
Normal itself. There is no closed-form expression for its distribution
and so direct maximum likelihood estimation is not feasible. Instead,
King reparametrizes the model so that the \(\vb B_g\) can be
analytically integrated out, leveraging that (a) the restriction of the
truncated bivariate Normal distribution to the tomography line is still
a (univariate) truncated Normal distribution, and (b) univariate
truncated Normal distributions have a closed-form CDF and normalizing
constant \autocite[Appendix D]{king1997solution}. King's original
computational proposal remains a good strategy for the 2×2 case, even as
more advanced generic Markov chain Monte Carlo (MCMC) methods have been
developed.

\begin{rrchange}

\subsection{Other parametric models}\label{other-parametric-models}

Other models impose distributional assumptions on different latent
quantities. \textcite{imai2008bayesian} extend King's fully parametric
framework by modeling the local parameters on a logit scale, allowing
more covariates and a flexible mixture of distributions, while
preserving the geography-level accounting identity.

The older Thomsen model \autocite{thomsen1987logit} may be best thought
of as estimating an individual IRT model with aggregate means. Voter's
latent utilities are distributed as bivariate Normal, and the method
uses proportion \(\bar{Y}\) and \(\bar{X}\) transformed by a probit
function to estimate the distribution's correlation parameters. It
constrains that distribution so that the observed \(\bar{X}\) or
\(\bar{Y}\) matches the CDF evaluated at 0 on each respective dimension.
This model does not leverage the accounting identity or the linear
structure of the problem. It also does not produce local estimates or
use covariates.

Yet other approaches cast estimation as constrained optimization. For
instance, \textcite{pavia2024improving} estimate global parameters that
minimize the discrepancy between local margins and fitted margins under
an L1 norm.\footnote{In our notation, they obtain a \(\hat{\vb{B}}\)
  that minimizes the sum of
  \(| \bar{Y}_g - \hat{\vb{B}}^\top\bar{\vb X}_g |\) across geographies.}
These approaches are analogous to parametric models in that they must
minimize some user-specified loss function, but differ in that this loss
function is not connected to a statistical model or estimand.

Finally, the neighborhood model \autocite{freedman1991ecological} allows
local parameters to arbitrarily vary across geographies, while assuming
them to be identical for all groups within a geography:
\begin{equation}\protect\phantomsection\label{eq-nbhd-model}{
\vb B_g = b_g\cdot\vb 1\ \text{for each}\ g, 
}\end{equation} where \(b_g\) is a scalar parameter and \(\vb 1\) is a
\(K\)-length vector of ones. Substitution of Eq.~\ref{eq-nbhd-model}
into the accounting identity immediately gives \(\bar Y_g\) as estimates
of \(b_g\). In essence, the neighborhood model replaces CCAR with its
own strong ignorability assumption. The model can also only hold at one
level of aggregation: if the assumption in Eq.~\ref{eq-nbhd-model} holds
at the precinct level, then it cannot generally also hold at the county
level \autocite{gelman2001models}.

\end{rrchange}

\subsection{Count models and beyond the 2×2
case}\label{count-models-and-beyond-the-22-case}

Researchers are interested in modeling discrete choice behavior across
more than two choices. However, in these higher dimensions, known as the
\emph{R×C} case, \textcite{king1997solution}'s sampling strategy faces
challenges because multivariate truncated Normal distributions involve
intractable normalizing constants.\footnote{Recent expectation
  propagation (EP) methods can quickly produce accurate approximations
  \autocite{cunningham2011gaussian}. Progress has also been made in
  sampling quickly from truncated multivariate Normal distributions,
  using elliptical slice sampling \autocite{wu2024fast}.}

\begin{rrchange}

Instead, researchers have turned to modeling the counts of individuals
in each geography. In the 2×2 case, the number of individuals with
\(Y = 1\) is the fraction \((B_{1} \bar{X}_1 + B_2 \bar{X}_2)\)
multiplied by the total population \(N,\) so the count approach models a
binomial count process with probability \(B \bar{X}\) and trials \(N\),
or \(\textrm{Binom}(N, B\bar{X})\). Initially proposed by
\textcite{brown1986aggregate} and developed for the 2×2 case in
\textcite{king1999binomial} and \textcite{wakefield2004ecological}, this
model can then be extended to a multinomial for more categories. Both
count models and models of fractions can be thought of as a linear
regression. We review two prominent models, \textcite{rosen2001bayesian}
and \textcite{greiner2009rxc}. The count models can extend to multiple
outcome choices, as we will show, but they also bring their own modeling
challenges.

\end{rrchange}

Formally, we generalize \(Y\) to be multivariate instead of binary by
letting it take on one of \(J\) levels, and let \(M_{g1},\dots,M_{gJ}\)
be the counts of the number of individuals in each level of \(Y\) in
geography \(g\). The parameter \(\vb B_g\) is now a matrix, with \(K\)
rows and \(J\) columns. For example, \(B_{g,\text{white},\text{dem}}\)
is the proportion of White voters in geography \(g\) who vote for the
Democratic candidate.

King's 1997 model is expressed in terms of the local finite-sample
parameters \(\vb B_g\), which must strictly satisfy the accounting
identity (Eq.~\ref{eq-acct-id}). The \emph{R×C} models take a different
approach, parametrizing the model in terms of \(\vbg\beta_g\), the
superpopulation counterpart to \(\vb B_g\).
\textcite{rosen2001bayesian}, in the widely used \texttt{eiPack} R
package, propose a simple count model for \(\vb M_g\):
\begin{equation}\protect\phantomsection\label{eq-rosen}{
 \underbrace{\vb M_g}_{[J\times 1]} \iid \mathrm{Multinom}(N_g,  \underbrace{\vbg\beta_g^\top\bar{\vb X}_g}_{[J\times 1]})\ \text{for each}\ g.
}\end{equation} Because \(\vbg\beta_g\) are superpopulation parameters,
which may be different from the unobserved \(\vb B_g\), they need not
satisfy the accounting identity. This additional wiggle room aids
greatly in estimation of the local parameters: the accounting identity
is satisfied only in expectation.

\begin{rrchange}

The choice of Multinomial distribution, however, limits the expressive
power of the model and imposes several substantive assumptions.
Eq.~\ref{eq-rosen} is appropriate when each outcome choice in a
geography \(g\) is a draw from a single \(J\)-length probability vector,
identically and independently from other individuals in the geography.
This sits uneasily with the possibility that outcomes vary with group
membership \(k\). For example, suppose \(K=2\) and all White voters in
precinct \(g\) prefer Republicans
(\(\beta_{g,\text{white},\text{dem}}=0\)) while an equally sized group
of Black voters all prefer Democrats
(\(\beta_{g,\text{black},\text{dem}}=1\)). Then Eq.~\ref{eq-rosen}
assumes that \emph{all} the voters in the precinct, White and Black,
vote for Democrats with the same probability of 0.5. While on average
this produces the correct number of Democratic votes in the precinct,
the model formally rules out both heterogeneity and correlation in
individuals' outcomes within precincts, even as it attempts to model
different outcomes by group.

\end{rrchange}

Further, \textcite{rosen2001bayesian} as well as
\textcite{brown1986aggregate} adopt a Dirichlet prior for each row
(group membership) of \(\vbg\beta_g\), with each Dirichlet being
independent. In the racial voting example, the probability a White voter
in geography \(g\) votes for a Democrat is estimated independently from
the probability that a Black voter in that same geography votes for a
Democrat. This is a marked departure from King's 1997 model, where this
pair of parameters are drawn from a bivariate distribution with an
estimated correlation parameter. The Dirichlet distribution (with \(J\)
parameters) is less flexible than a Normal distribution (with
\(J + J(J-1)/2\) parameters). Specifically, under Dirichlet, the
correlation between any two entries is pre-determined by the mean of
each entry and the overall dispersion of the distribution, whereas the
correlation in a multivariate Normal can be freely specified.

\textcite{greiner2009rxc} directly addressed these drawbacks of the
\textcite{rosen2001bayesian} model by proposing a more flexible model
which allows for more correlation within \(\vbg\beta_g\). They use a
slightly different count model which models the local counts
\(N_{gk}\vb B_{gk}\) for each category \(k\) individually rather than
the totals across categories \(\vb M_g\): \[
\underbrace{N_{gk}\vb B_{gk}}_{[J\times 1]} \iid \mathrm{Multinom}(N_{gk}, \vbg\beta_{gk})\ \text{for each}\ g,k,
\] where \(\vb B_{gk}\) and \(\vbg\beta_{gk}\) are now \([J\times 1]\)
vectors describing the conditional outcomes for individuals in category
\(k\) and geography \(g\). This setup at least reflects that voters in
different racial groups vote differently.\footnote{A parallel in the 2×2
  case is \textcite{wakefield2004ecological}'s contribution over that of
  \textcite{king1999binomial}'s 2×2 setup.} Further, to allow for more
correlation across each row of \(\vbg\beta_{g}\),
\textcite{greiner2009rxc} transform each row \(\vbg\beta_{gk}\) to a
vector of \(J-1\) logits, picking one of the \(J\) choices as the
reference category. These logit preferences are then modeled as a
multivariate Normal distribution, which allows for correlation across
the \(K\) categories. Unfortunately, their R implementation is no longer
publicly available on CRAN.

\begin{rrchange}

Despite these differences in model, the CEF of the outcome is still
linear for both R×C models, just as it is for King's model and Goodman's
regression. For both models, we have \[
\E[\vb M_g \mid N_g, \bar{\vb X}_g] = N_g\vbg\beta_g^\top\bar{\vb X}_g.
\] Thus both models can still be viewed as (multivariate) linear
regressions, with residuals distributed as a discrete mixture of
Multinomials.

\end{rrchange}

\subsection{Semiparametric modeling}\label{sec-semipara}

\begin{rrchange}

While past ecological inference methods do allow users to incorporate
covariates, all of them impose specific functional forms and are not
especially concerned about researcher degrees of freedom.\footnote{Both
  \textcite{king1997solution} and \texttt{eiPack}'s implementation of
  \textcite{rosen2001bayesian} allow \(\vbg\beta_g\) to be further
  modeled as a linear function of user-specified covariates, but allow
  no other functional form. As of writing, \texttt{eiPack} only accepts
  one covariate, and \textcite{king1997solution} cannot generate
  estimates when the covariate matrix is rank deficient.
  \textcite{hanushek1974model} include covariates as controls but do not
  interact them as Eq.~\ref{eq-cef-lin} suggests. Modeling the
  coefficients as a linear function of the \(\bar{X}\) variables
  themselves has also been attempted, and is referred to as the linear
  contextual model \autocite{blalock1984contextual}.} We conclude this
section by considering an alternative \emph{semiparametric} approach
that makes no parametric distribution assumptions about the error term
but still estimates the global estimand correctly, as long as CAR holds.
Theoretical developments in semiparametric statistics apply well to
ecological inference because of the accounting identity.

\end{rrchange}

Recall that Eq.~\ref{eq-cef}, reproduced below, always holds once CAR is
satisfied with covariates \(\vb Z\): \[
\E[\bar Y_g \mid \vb Z_g, \bar{\vb X}_g] = f(\vb Z_g)^\top \bar{\vb X}_{g}.
\] Thus, if \(f(\vb Z_g)\) could be estimated, then following
Proposition~\ref{prp-id}, we could estimate \(\vbg\beta\) without any
additional assumptions. We now relax the linearity-on-covariates
assumption by expanding \(\vb Z_g\) into a rich basis \(\Phi(\vb Z_g)\),
such as splines, interactions, or trees.\footnote{The \texttt{bases}
  package \autocite{bases} provides a convenient way to incorporate such
  preprocessing steps within the ecological inference procedure.}
Well-developed statistical theory establishes that if the basis
expansion \(\Phi\) is rich enough, and grows richer as the sample size
increases, then \(f(\vb Z_g)\) can be consistently estimated
\emph{nonparametrically} \autocite[see, e.g.][]{shen1994convergence}.

In practice, this involves picking a specific basis expansion \(\Phi\),
interacting the expanded \(\Phi(\vb Z_g)\) with \(\bar{\vb X}_g\), as in
Eq.~\ref{eq-cef-lin}, and then regressing \(\bar Y_g\) on these terms.
Linear-in-covariates is then one out of many possible functional forms.
The large number of terms in the basis expansion \(\Phi\) means that
ordinary least squares will likely overfit, or be unable to be fit at
all. A penalized regression is therefore recommended. The value of the
penalty can be selected by leave-one-out cross-validation, for which a
closed-form expression exists for ridge regression.

\begin{rrchange}

To achieve the best statistical properties, one additional ingredient is
needed: the so-called \emph{Riesz representer} for \(\vbg\beta\). For
our purposes here, the Riesz representer for racial group \(k\) is a set
of weights, one for each geography \(g\), such that weighted averages of
\(\bar Y_g\) with these weights can estimate \(\vbg\beta\). It can be
thought of as a generalization of the inverse propensity score model for
the treatment (here, \(\bar{X}_k\)) in causal inference. Just like the
augmented inverse propensity weighting (AIPW) estimator from causal
inference, we can combine the Riesz representer weights with the fitted
regression function to produce a statistically improved estimate of
\(\vbg\beta\). This combination is referred to as \emph{double/debiased
machine learning} \autocite[DML,][]{chernozhukov2018double}.\footnote{
  The use of the Riesz representer is motivated by the following
  challenge. When there are many terms in \(\Phi(\vb Z_g)\), the ridge
  penalty in the regression will be large, and the resulting
  regularization bias will lead to bias in the primary estimate of
  \(\vbg\beta\). This is a well-known problem when estimating a
  high-dimensional nuisance function (i.e., a function that is not the
  main parameter of interest, here, \(f\)). DML methods address this by
  learning a \emph{second} nuisance function known as the Riesz
  representer. The DML estimator that combines the two functions is a
  generalization of the AIPW estimator \autocite{robins1995analysis}.
  Like AIPW, DML is doubly robust, and so tolerates misspecification of
  either nuisance function, as long as the other is correctly specified.}

\end{rrchange}

We briefly summarize the approach of \textcite{eimethods}, which
implements this approach in the R package \texttt{seine} (standing for
\ul{s}emiparametric \ul{e}cological \ul{in}ferenc\ul{e}):

\noindent \textbf{Procedure 2}

\begin{enumerate}
\def\labelenumi{\arabic{enumi}.}
\tightlist
\item
  Fit the linear regression model of \(\bar Y\) on expanded covariate
  bases \(\Phi(\vb Z_g)\) interacted with \(\bar{\vb X}_g\).
\item
  Calculate the Riesz representer weights \(\alpha_k\) for each racial
  group \(k\) \autocite{eimethods}.
\item
  Plug in \(\bar X_{gk}=1\) (and \(\bar X_{gk'}=0\) for all
  \(k'\neq k\)) into the fitted regression model, which produces
  predictions \(\hat f_k(\vb Z_g)\).
\item
  Augment the fitted regressions with the Riesz representer to form the
  \emph{score} for each geography \(g\) and category \(k\),

  \begin{rrchange}

   $$
   s_{gk} = \underbrace{\hat f_k(\vb Z_g)\,|G|\,\overbrace{\frac{N_{gk}}{N_k}}^{\substack{\text{$g$'s share}\\\text{of group $k$}}}}_{\text{imputation}} + \underbrace{\hat\alpha_k\,(\bar Y_g - \hat{\bar Y}_g)}_{\text{debiasing correction}},
   $$
   where $|G|$ is the number of geographies and $\hat{\bar Y}_g:=\sum_{k} \hat f_k(\vb Z_g)\bar X_{gk}$ is the outcome average implied from the fitted model.

   \end{rrchange}
\item
  Use the mean of the scores \(s_{gk}\) across geographies as the
  estimate of \(\beta_k\). The standard deviation of the scores divided
  by the square root of the number of geographies serves as its standard
  error.
\end{enumerate}

\begin{rrchange}

This approach has several advantages over EI methods that do not include
covariates or only include covariates through limited functional forms.
It is consistent under much weaker assumptions and has the smallest
possible variance as the amount of data increases. By construction, the
regression respects the partially linear form of Eq.~\ref{eq-cef}, since
\(\bar{\vb X}_g\) is interacted with \(\Phi(\vb Z_g)\), while allowing
for nonlinearities through the analyst's use of a basis expansion.
Computationally, it is fast to fit, because both the regression for
\(f\) and the estimator for the Riesz representer \(\alpha\) have
closed-form solutions, and, unlike the early DML literature, we need not
estimate these two components using held-out split samples of the data
\autocite{chen2022debiased}. When outcome \(Y_i\) is binary or
categorical, bounds can be imposed at step 1 to improve the efficiency
of the regression, although the final debiased estimate need not itself
satisfy those bounds. \texttt{seine} also provides estimates for the
local parameters, using the Duncan--Davis bounds under additional
assumptions.

\end{rrchange}

\section{Empirical Validation}\label{sec-empirics}

We next turn to two empirical applications of these ecological inference
methods, which illustrate the challenge of meeting the CAR condition.
While violations of the assumption are not observed in practical
applications, in rare cases, the ground truth estimand is observed. We
discuss two common applications in political science, each organized by
the patterns in the observable data, the performance of methods in
estimating the unobserved quantities of interest, and patterns of
confounding.

\subsection{Partisanship and vote choice by
race}\label{partisanship-and-vote-choice-by-race}

In the most common application of ecological inference, the outcome
\(Y\) is an individual's vote choice for a candidate or party, and \(X\)
is the race/ethnic membership of the voter
\autocite{greiner2010exit,freedman1991ecological,kuriwaki2024geography}.
When a White majority votes predominantly for Republicans while a
sizable Black and Hispanic population votes predominantly for Democrats,
for example, U.S. jurisprudence of the Voting Rights Act has recognized
a need for states to redraw their districts in a way that essentially
lets Democrats win office. Determining if these conditions are met is an
ecological inference problem, because election reporting units contain a
mix of White and non-White voters. While past work has examined
ecological inference methods on these problems (see Appendix D), none
have considered how including covariates affects predictive accuracy.

Here we use North Carolina voters' party registration on the voter file
as a proxy for their partisan voting behavior, and evaluate how well key
ecological inference methods can recover party registration by race.

\begin{rrchange}

Voter files in North Carolina are public, individual-level
administrative datasets that include a registrant's self-identified
racial group, party registration, and precinct. Although party
registration is not vote choice, the two are heavily correlated
\autocite{kuriwaki2024geography}.

\end{rrchange}

\begin{figure}

\begin{minipage}[c]{0.50\linewidth}

\centering{

\pandocbounded{\includegraphics[keepaspectratio]{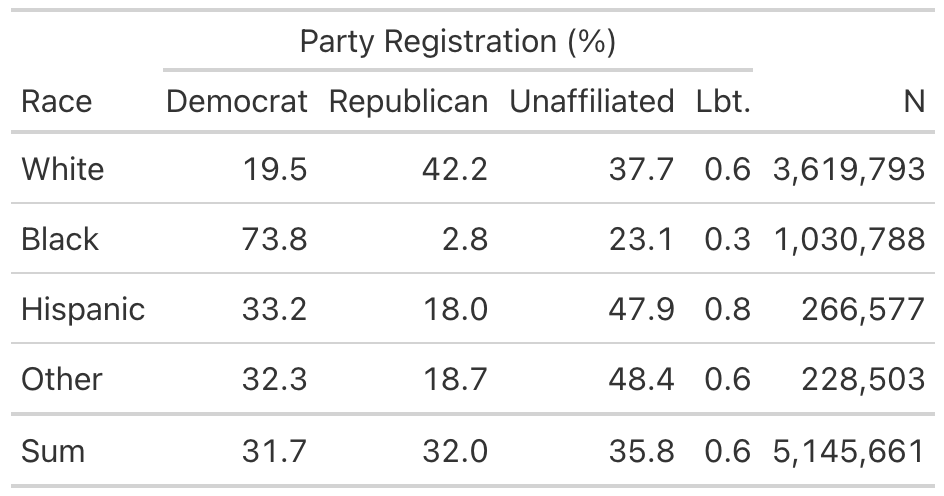}}

}

\subcaption{\label{fig-NC1-a}Ground Truth Joint Distribution}

\end{minipage}%
\begin{minipage}[c]{0.50\linewidth}

\centering{

\pandocbounded{\includegraphics[keepaspectratio]{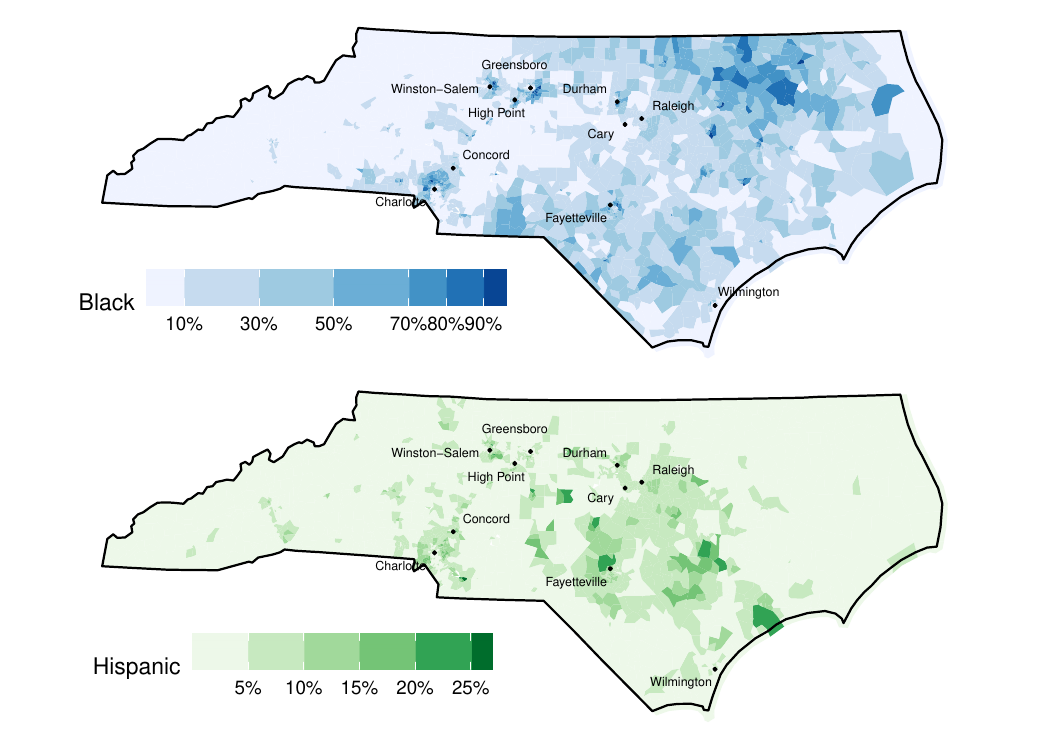}}

}

\subcaption{\label{fig-NC1-b}Geography of Racial Minorities}

\end{minipage}%

\caption{\label{fig-nc1}\textbf{The Distribution of Party and Race
Registration in North Carolina.} We use the voter file as a proxy
validation for measuring racially polarized voting. Panel (a) shows the
joint distribution of party and race as recorded in the state's public
voter file. Cells show row percentages within each race, and are the
quantities of interest. Panel (b) shows the geographic distribution of
the two racial minorities. Black registrants live in the Northeast part
of the state (Piedmont) and around large cities. Hispanic registrants
constitute only 5\% of the entire dataset and are concentrated in rural
parts of the state.}

\end{figure}%

The North Carolina dataset covers 2,465 precincts, with each precinct
averaging about 1,600 voters (See Section~\ref{sec-app-data} for details
on data construction).\footnote{Although some of these include
  non-voting active registrants, we refer to all individuals as voters
  for simplicity.} The state is evenly divided between registered
Democrats (31\%), Republicans (31\%), and non-affiliated voters, but
racial groups sort into distinct party patterns. Only about 19\% of
White voters register as Democrats, while 74\% of Black voters do the
same (Figure~\ref{fig-NC1-a}).

The CCAR condition is violated if voters of the same racial group lean
toward different parties depending on how prevalent their group is in
the local geography. Black voters are about 20\% of the dataset, and
concentrated around a few of the major cities---Greensboro and
Charlotte---but especially concentrated in the Black Belt of the
northeast coastal plains (Figure~\ref{fig-NC1-b}). A few precincts are
almost completely composed of Black voters: 3 out of the 2,465 precincts
are over 95\% Black and 17 are over 90\% Black. Hispanic voters are only
5\% statewide, but also nonrandomly dispersed. No precinct is more than
27\% Hispanic, and about 200 precincts are between 10\% and 20\%
Hispanic. They tend to be concentrated in suburbs around the center of
the state, and residential patterns are correlated with manufacturing,
agriculture, and food processing (Figure~\ref{fig-NC1-b}).

We test four methods on this data: (1) the semiparametric estimator
\texttt{seine} with covariates entering through a tensor-product sieve
expansion, (2) an OLS regression with no covariates (i.e., Goodman
regression), (3) Rosen et al.'s count model, as implemented in
\texttt{eiPack} \autocite{lau2007eipack}, with no covariates, modeling
the 4-by-4 matrix at once, and (4) King's 2×2 method applied to each
racial group and each outcome separately. The covariates we included in
\texttt{seine} were precinct level education, age, and income as
measured from the American Community Survey, urbanicity as measured by
distance to a major city or university, and population density.

\begin{figure}

\centering{

\pandocbounded{\includegraphics[keepaspectratio]{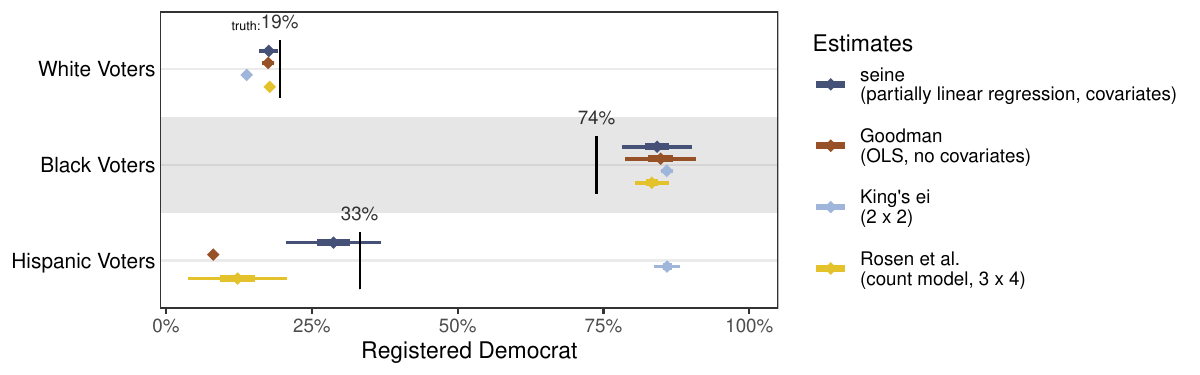}}

}

\caption{\label{fig-nc2}\textbf{Accuracy of EI Methods in Uncovering
Partisanship among Racial Groups.} Ecological inference estimates for
the percentage of each racial group that is registered for the
Democratic Party, with 95\% and 50\% confidence intervals. The true
values of the registrations are shown in gray vertical lines and
labelled. Estimates from different methods are shown with the black
line. \texttt{seine} includes the covariates education, median income,
median age, density, and distance to a city/university. Rosen et al.'s
model refers to the multinomial Dirichlet model available in
\texttt{eiPack}. Estimates for other outcomes are shown in table form in
Table~\ref{tbl-nc2-full}.}

\end{figure}%

The resulting estimates appear in Figure~\ref{fig-nc2} with the ground
truth as a reference point. \texttt{seine} estimates \(17.5\)\% of White
voters are registered Democrats with a 95\% CI of \([0.159, 0.191]\),
where the true value is \(19.5\)\%, and it estimates that \(44\)\% of
White voters are registered Republicans with a 95\% CI of
\([0.422, 0.469]\), where the true value is \(42.2\)\%. Neither the
Goodman regression nor King's EI method have confidence intervals that
cover the truth. Their point estimates are farther from the true values,
and their intervals are narrower. Among Hispanic voters, shown in the
bottom row, conventional methods also have worse estimates. Goodman
underestimates the Democratic percentage among Hispanics; the King model
estimates overestimate it by more. This is an unusual case where the
Goodman regression and the King model give opposite answers.
\texttt{seine} correctly estimates the Hispanic degree of party
registration.

Black voters have more stubborn estimation challenges. All methods
overestimate the Democratic percentage among Black voters by about
10--20 points. The \texttt{seine} methods come the closest but the
confidence interval does not cover the true values. The overestimation
of Democratic leaning among Black voters also goes hand-in-hand with the
underestimation of the Republican leaning of Black voters. The linear
regression-based model gives negative estimates of \(-2.0\)\% and
\(-1.6\)\% (Table~\ref{tbl-nc2-full}). King and Rosen's models are
advantaged by constraining estimates to be valid proportions between 0
and 1.

Concerningly, all models miss in the direction of \emph{overestimating}
White vs.~Black racial polarization. In truth, there is a 54 point gap
in Democratic party registration between White voters (19.5\% registered
Democrat) vs.~Black voters (73.8\%, see Figure~\ref{fig-nc1} (a)).
However, King's model estimates the White-Black gap to be 72 points. It
makes that mistake by underestimating the Democratic registration of
White voters by six points and overestimating the Democratic
registration of Black voters by twelve. Adding covariates
(\texttt{seine}) reduces this estimation error by 6 percentage points,
but the implied racial gap is still an overestimate.

Finally, Rosen et al.'s count model produces biased estimates similar to
linear regression, but suffers from an additional challenge: Its
estimates are highly sensitive to the presence of rare outcome
categories in a way that other methods are not. The count model's
estimates are similar to the others only when Libertarian votes are
grouped together with one of the three larger groups before estimation.
When the small Libertarian outcome choice (comprising 0.6\% of the
population) is treated as a fourth category instead of being lumped
together into the third, its estimates for other outcome groups change,
for example missing the estimate of Black voters' Democratic
registration by over 40 points (Appendix
Table~\ref{tbl-nc-rosen-sensitivity}).

\begin{figure}

\centering{

\includegraphics[width=0.8\linewidth,height=\textheight,keepaspectratio]{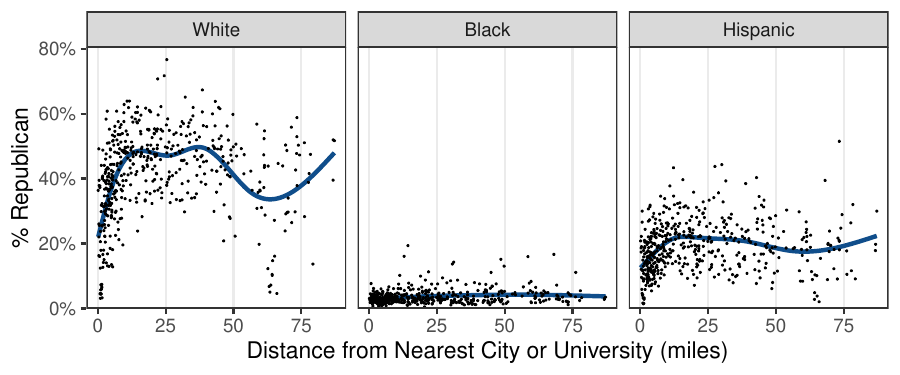}

}

\caption{\label{fig-nc3}\textbf{Potential Confounders for Racially
Polarized Voting.} Each point is a precinct, sorted by distance to a
nearest city or university, a potential confounder, on the horizontal
axis. The vertical axis shows the ground truth level of Republican
registration in those racial groups in those precincts (the local
parameter \(b\) of interest). A GAM regression showing the line of best
fit is shown in blue. A sample of 500 precincts with at least 20 voters
in all three racial groups is shown for clarity.}

\end{figure}%

What sort of confounding does the inclusion of covariates help solve?
One potential confounder is the urbanicity of an area. Urbanicity is
clearly correlated with the prevalence of a racial group. If urbanicity
is also correlated with how Democratic the White voters in that region
are, for example, estimates will be biased. In Figure~\ref{fig-nc3}, we
operationalize urbanicity as the distance from the center of a large
city \autocite[as in][]{rodden2019cities} or a R1 university, and show
its relationship with the parameter of interest. The Republican
preference of racial groups indeed differs systematically by the
urbanicity of the precinct. White and Hispanic voters in the urban core
are systematically less Republican than in distant areas. Moreover, the
relationship is nonlinear in the distance metric. This suggests that
allowing covariates to enter the model in a flexible functional form as
we do with \texttt{seine}'s bases is important. The limited impact of
the covariate in Figure~\ref{fig-nc2} may be due to the limited
variability that exists in racial composition once urbanicity is
controlled for.

\subsection{Ticket splitting}\label{sec-ticket-splitting}

In studies of ticket splitting the outcome \(Y\) is a vote choice in one
office, and \(X\) is the vote in another. A voter who votes for party
\emph{A}'s candidate in one office but votes for party \emph{B}'s
candidate in another office is called a ticket splitter. The prevalence
of these ticket splitters shapes the degree of divided government and
measure the degree of nationalized partisan behavior. Each geographic
unit provides the voteshares of candidates in their respective contests
separately.

We use a tranche of anonymous ballot records (\emph{cast vote records})
to observe the ground truth rates of ticket splitting exactly. This
dataset from 2020 \autocite{kuriwaki2024cast} records a ballot's vote
choice for every contest on a ballot, including President and U.S.
House. It also records the precinct in which the vote was cast. We seek
to estimate ticket-splitting rates for U.S. House candidates in 63
congressional districts.

\begin{rrchange}

In this dataset, 5.2\% of those who voted for a major-party presidential
candidate split their ticket in the U.S. House race for the opposite
party, and 3.9\% of them abstained or voted third party.\footnote{Using
  the covariates we later discuss, we can project our estimates to a
  hypothetical national population. Across all 435 congressional
  districts, we estimate a ticket splitting rate of 5.9\% and an
  abstention/third-party rate of 4.0\%.} Our validation of EI for ticket
splitting is the most comprehensive to date (see
Section~\ref{sec-app-data}).

\end{rrchange}

\begin{figure}

\begin{minipage}[c]{0.33\linewidth}

\centering{

\includegraphics[width=1\linewidth,height=\textheight,keepaspectratio]{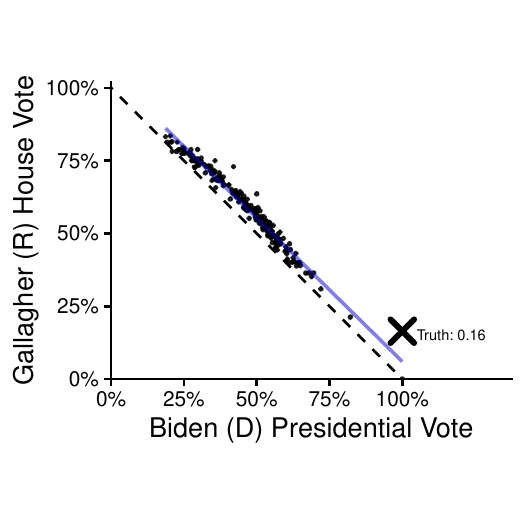}

}

\subcaption{\label{fig-cvr-d}Precinct values in WI-08}

\end{minipage}%
\begin{minipage}[c]{0.67\linewidth}

\centering{

\includegraphics[width=1\linewidth,height=\textheight,keepaspectratio]{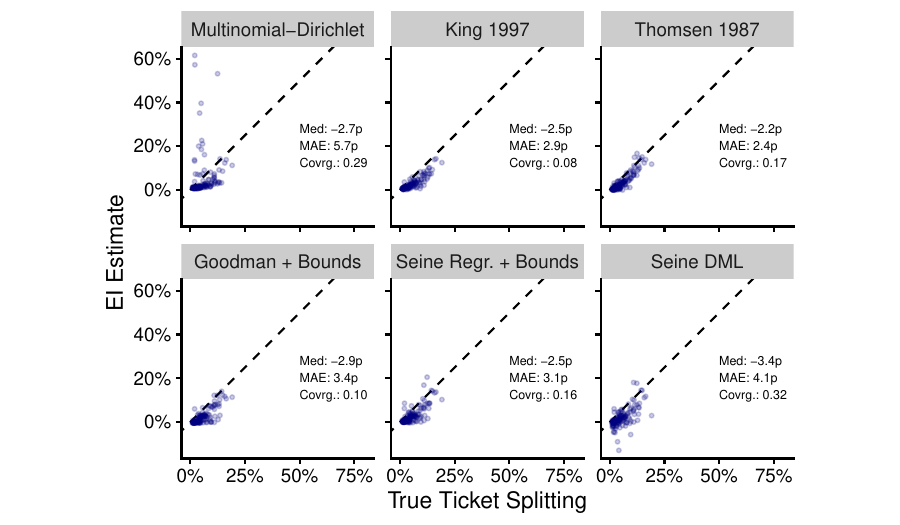}

}

\subcaption{\label{fig-cvr-estimates}Validation of Global Estimates}

\end{minipage}%

\caption{\label{fig-cvr-2}\textbf{Predictive Performance of Ticket
Splitting.} (a) Example of precinct-level data in Wisconsin's 8th
congressional district. Dotted line indicates a 45 degree line, and blue
line is the OLS best fit. Cross indicates the true percentage of Biden
voters who split their ticket for Gallagher. (b) Each facet shows
estimates of congressional district (CD)-level ticket splitting for a
given method. All facets use the same n = 63 CDs. Estimates in the first
rows come from \textcite{rosen2001bayesian},
\textcite{king1997solution}, and \textcite{thomsen1987logit}.
\begingroup\color{RevisionColor}The second row comes from the \texttt{seine} software.\endgroup \\
Avg. is mean error, MAE is mean absolute error, and Covrg. is empirical
coverage of the estimated 95\% confidence intervals.}

\end{figure}

As one example, consider the ballots from a part of Wisconsin's 8th
congressional district, where 16\% of those who voted for Joe Biden
voted for the Republican incumbent Michael Gallagher.\footnote{This may
  be due to Gallagher's name recognition as an incumbent, his candidate
  quality, or his moderate stance relative to Donald Trump
  \autocite{kuriwaki2023ticket}.} Figure~\ref{fig-cvr-2}(a) shows the
observable, aggregate voteshares at the precinct level in this district.
The incumbent's overperformance appears constant across all levels of
the presidential Democratic lean in the precinct. The best fit line
evaluated at \(\bar{X} = 1\) is at 9\%, which is a substantial
underestimate of the true ticket splitting rate of 16\%. The line
evaluated at \(\bar{X} = 0\) implies an impossible value: that 102\% of
Trump voters voted for Gallagher.

We find that the methods tend to underestimate the degree of ticket
splitting, or, in other words, overestimate the degree of partisan
congruence in the two offices. Figure~\ref{fig-cvr-2}(b) compares the
estimates on the vertical axis with the CD-level ground truth on the
horizontal axis. The first Multinomial Dirichlet model by
\textcite{rosen2001bayesian} estimates ticket splitting rates that are
too low, with over half of the estimates being underestimated by 2 to 3
points.\footnote{ As in our results for North Carolina, differentiating
  between 4 outcome categories makes the errors even worse than the
  binarized version shown here.} King's 1997 model in the second panel
generates estimates that have a lower absolute error than the count
models. Most estimates still underestimate the true levels of ticket
splitting by the same amount.

\begin{rrchange}

The Thomsen model, shown in the third facet, does quite well, achieving
the lowest mean absolute error. This is consistent with a similar
investigation by \textcite{park2014ecological}. The model's
transformation of input data to the probit scale is consequential for
data near the extremes. We have found that the Thomsen model tends to do
well when the global parameters are small. However, this method is
restricted to the 2×2 case, does not support covariates, and does not
produce local estimates. It too tends to underestimate ticket splitting.

\end{rrchange}

The linear regression yields similar, if somewhat noisier, patterns. In
panels four and five, we estimated the regression component by
specifying the ridge regression to estimate a solution that is strictly
within \([0, 1]\).\footnote{Misspecifications in flexibly modeling
  \(f(\vb Z_g)\) could produce implausible estimates that are out of the
  possible range of \(Y\). In these cases, the \texttt{seine} software
  uses quadratic programming which enforces that the predicted values of
  the regression when each \(\bar X_{gk}\) is set to \(1\) must lie
  within the range of \(Y\) (e.g., 0 to 1).} The models thus avoid
producing negative, infeasible estimates that linear regression is prone
to give in this case. Without covariates, the mean absolute error is
\(3.4\) points, with estimates again underestimating ticket splitting by
the same amount. Including covariates in the regression improves
estimates modestly,\footnote{We include the following covariates:
  county-level median income, county-level proportion White, county
  population density, county-level proportion of elderly residents, and
  indicators for five equally populated bins of the precinct-level
  presidential vote. The limited impact of covariates we see in this
  example may be due to the fact that the covariates we could collect
  (county-level) were much coarser than the geographic units
  (precinct-level).} to comparable error magnitudes as the King model.
Finally, while the DML estimate is not bounded and thus has higher
error, its estimated standard errors have the coverage rate closest to
nominal coverage among the six examples tested.

\begin{figure}

\centering{

\includegraphics[width=0.6\linewidth,height=\textheight,keepaspectratio]{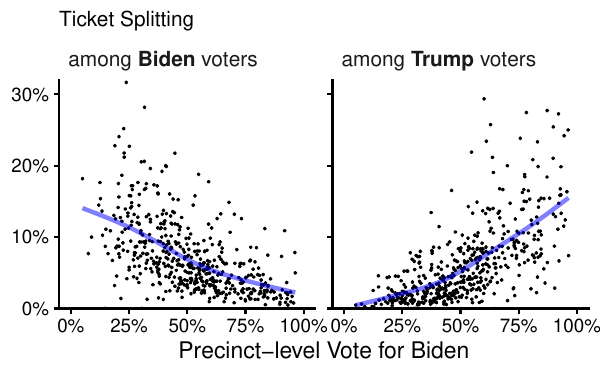}

}

\caption{\label{fig-cvr-3}\textbf{Variation in Ticket Splitting Rates.}
The relationship between \(\bar{X}\) and quantities of interest across a
random sample of 500 precincts with more than 100 voters for visual
clarity. Each point is a sampled precinct, sorted on the horizontal axis
by the percentage of total votes for the Democratic presidential
candidate. Local estimands \(B_{g}\) are heavily correlated with
\(\bar{X}\).}

\end{figure}%

A particular form of CAR violation appears to explain the
underestimation of ticket splitting. In Figure~\ref{fig-cvr-3}, we plot
how the degree of ticket splitting in each partisan group varies by the
partisan lean of the precinct. Biden voters are least likely to split
their ticket when they are in nearly unanimous Biden precincts (left
panel). Similarly, Trump voters are least likely to split their ticket
when they are in nearly unanimous Trump precincts (right panel).

\begin{rrchange}

Put together, voters in precisely the precincts that have
disproportionate leverage in the regression are systematically different
from other voters, in a way that is prone to underestimation of ticket
splitting. Furthermore, because the confounding is due to the group
composition itself, controlling for relevant confounders leaves little
variation to identify the parameters.

\end{rrchange}

\subsection{Sensitivity to
Confounders}\label{sensitivity-to-confounders}

\begin{rrchange}

In practical applications, there is no ground truth to validate against.
Practitioners cannot tell how much the use of covariates actually
improves the estimates. Further, although the semiparametric estimator
removes some researcher degrees of freedom, researchers still choose
which basis expansion to use. And fundamentally, these methods can only
use covariates that are observed by the researcher. In short, CAR is an
untestable assumption. The sensitivity analysis developed in new work
can address these concerns, and applying it to the three examples in the
paper is instructive.

Suppose that CAR holds conditional on observed covariates \(Z\) and an
unobserved variable \(U\). We denote the estimand for category \(k\)
when conditioning under both \(Z\) and \(U\) as \(\beta_k,\) and denote
the (incorrect) result in the \emph{short} regression where the
unobserved \(U\) is not included as \(\beta^{\textrm{short}}\).
\textcite{eimethods}, extending work from
\textcite{chernozhukov2022long}, show that the bias from the short
regression, \(\textrm{bias} := \beta^{\textrm{short}}_k - \beta_k,\) has
magnitude bounded by:
\begin{equation}\protect\phantomsection\label{eq-bias-bound}{
|\textrm{bias}| \leq \rho\, S\, C_\gamma\, C_\alpha,
}\end{equation} where \(\rho\) is a correlation that has a maximum value
of 1, \(S\) is a scaling factor that can be estimated by observed data,
and the two variables \(C\) are confounding measures that are
interpretable as the relevance of the unobserved variable:\footnote{See
  \textcite{eimethods} for a more detailed discussion of the
  interpretation of \(C_\alpha^2\).}
\begin{equation}\protect\phantomsection\label{eq-bias-C-def}{
C_\gamma^2 = R^2_{\bar Y \sim U \mid \bar{\vb X}, \vb Z}
\qand
C_\alpha^2 \approx \text{increasing in }R^2_{\bar{\vb X} \sim U \mid \vb Z}.
}\end{equation}

Each \(R^2\) above is a partial \(R^2\) of linear regressions using our
variables.\footnote{The partial \(R^2_{y\ \sim\ u\mid x}\) is computed
  by first regressing \(y\) and \(u\) on \(x\) and taking residuals, and
  then calculating the standard \(R^2\) of the residuals of \(y\) on the
  residuals of \(u\).} That is, the worst case absolute bias of the
observable estimate can be bounded under scenarios of hypothetical
confounding between the confounder and the outcome (\(C_\gamma\)) and
the confounder and the predictor (\(C_\alpha\)). We can suggest
reasonable values of these quantities by benchmarking with values that
replace \(U\) with observed variables.

\begin{figure}

\centering{

\includegraphics[width=1\linewidth,height=\textheight,keepaspectratio]{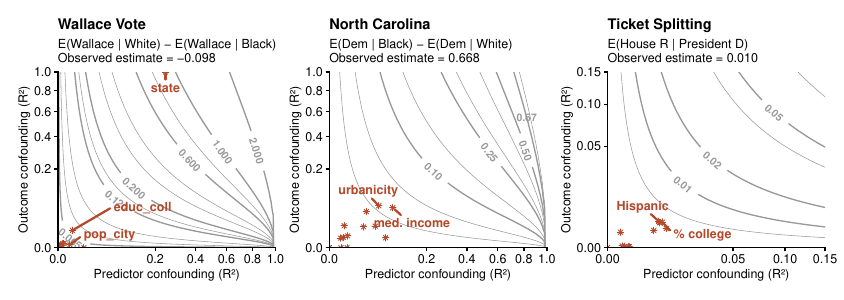}

}

\caption{\label{fig-sens}\textbf{Evaluating Sensitivity to Unobserved
Confounding.} Contour lines show the absolute value of the bias that an
unobserved confounder could induce in the point estimate as a function
of its confounding with the outcome (y-axis) and conditioning predictor
variable (x-axis). Axes are on the square-root scale. Each panel is
titled by the quantity of interest and its point estimate using the
observed covariates. See appendix for data and implementation details.}

\end{figure}%

Figure~\ref{fig-sens} plots the resulting biases. Each panel is best
read in three steps: start with the point estimate given by the
regression, find a value of the contour lines that would reverse the
estimate to a substantial amount, then ask whether such an unobserved
confounder is plausible given the benchmarked observed covariates.

The first example of the Wallace vote differential between White and
non-Whites has the wrong sign (\(-0.098\)). Any combination of the
confounding values \(C\) that leads to a contour value of about that
magnitude could, in other words, mask a true null. To assess whether a
bias of such magnitude could plausibly occur, we look to the associated
confounding of the existing covariates, shown as orange points. The
variable \texttt{state} has the highest associated bias of nearly 1,
meaning that omitting that variable could induce a bias of nearly 1 on
the --1 to 1 scale. This is consistent with the large state-by-state
heterogeneity we see in the raw data of Figure~\ref{fig-wal}.

The other panels lead to different conclusions. In the middle North
Carolina panel, the point estimate is of the correct sign, but
overestimated by about 10 points. Leaving out each of the existing
covariates, such as urbanicity and income, only leads to additional
biases of less than 0.10. It is hard for us to think of a covariate not
already in the model that would be strong enough to overturn the result
to zero, and so we remain confident in the finding of racially polarized
registration. In the right panel, the estimate for ticket splitting is
only a single percentage point. The benchmark covariate's associated
bias is small, but so are the estimates themselves. Here we conclude
that controlling for other covariates could reasonably affect the
estimate.

\end{rrchange}

\section{Conclusion}\label{conclusion}

This paper offers a perspective to understand the problem of inferring
conditional means from aggregate data, a technique known as ecological
inference. Our framework shows how all ecological inference methods for
point identification are fundamentally similar in their use of
regression. Existing literature on ecological inference often treats
linear regression as a simplistic approximation to a more complicated
data structure. But that may have obscured the fundamental problem of
controlling for confounders. Unless these are controlled for, ecological
inference of any sort introduces bias. The recent literature has also
underemphasized how the fact of linearity in aggregate data aids
estimation of models with covariates, which in turn invites recent
innovations in semiparametric models.

Using this framework, we evaluated two applications of ecological
inference in political science where ground truth is observed. EI
estimates of racial differences in partisanship tend to be overestimated
because racial minorities are typically more Democratic-leaning in
concentrated areas than in diverse ones. EI estimates of ticket
splitting tend to be underestimated in recent elections because the
correlation of voteshares between offices is strong, and a party's
supporters tend to split their ticket less in concentrated areas than in
diverse ones. In the first example, inclusion of covariates reduces the
impact of ecological fallacies, while improvements in point estimates
are less clear in the second. The intuition and challenges of
observational causal inference, such as controlling for confounders
while retaining enough variation in the residualized treatment for
stable estimation, carry on to ecological inference.

\begin{rrchange}

In future research, more covariates can be brought to specific empirical
problems, possibly from multiple levels of aggregation where the
covariates and group membership are measured at the individual level but
outcomes are measured at the aggregate level, or in cases where
individual survey data is available to assist modeling. Sensitivity
analyses can be used to understand and make more credible ecological
inferences. In studies of geographies, modeling the residential sorting
decisions of individuals may be fruitful, as it also can open a
framework for design-based inference \autocite{abadie2020sampling}.
Ecological inference is a challenging missing data problem, but our
reassessment points to several promising directions.

\end{rrchange}

\printbibliography[title=References]
\end{refsection}

\newpage{}

\begin{refsection}

\appendix

\renewcommand\thefigure{\thesection.\arabic{figure}}

\setcounter{figure}{0}

\section{Methodological Developments in Ecological
Inference}\label{sec-app-rev}

We summarize the proposals made in the literature for ecological
inference. The list below captures, to our knowledge, most of the major
work proposing a new method or methodological adjustment to ecological
inference. We omit work whose main purpose is to evaluate the numerical
performance of an existing method.

\begin{rrchange}

All of the work cited in this appendix, with the possible exception of
the Thomsen model, leverages the accounting identity. For example, the
identity appears as the tomography line in King's framework, as linear
constraints in the optimization approach, and as the source of the sharp
bounds in \textcite{duncan1953alternative}.

\end{rrchange}

\subsection*{Ecological Inference before
1997}\label{ecological-inference-before-1997}
\addcontentsline{toc}{subsection}{Ecological Inference before 1997}

\begin{enumerate}
\def\labelenumi{(\arabic{enumi})}
\tightlist
\item
  Classical foundations: \textcite{robinson1950ecological},
  \textcite{goodman1953ecological}, \textcite{goodman1959some}
\item
  Summary of literature and applications to political science:
  \textcite{achen1995cross}
\item
  Inclusion of control variables: \textcite{hanushek1974model}
\item
  Neighborhood model: \textcite{freedman1991ecological}
\item
  Latent utility model: \textcite{thomsen1987logit}
\item
  Multinomial models for (\emph{R×C}) elections:
  \textcite{brown1986aggregate}
\end{enumerate}

\begin{rrchange}

\emph{Comment}: Our article discusses these foundational papers.
\textcite{cleave1995evaluation} is a useful review of this literature.
They provide a validation exercise comparing Goodman regression, Brown
and Payne multinomial count models, the Thomsen latent utility model,
and raking with survey data.

\end{rrchange}

\subsection*{\texorpdfstring{Literature adjacent to
\textcite{king1997solution}, in political science and
statistics}{Literature adjacent to @king1997solution, in political science and statistics}}\label{literature-adjacent-to-king1997solution-in-political-science-and-statistics}
\addcontentsline{toc}{subsection}{Literature adjacent to
\textcite{king1997solution}, in political science and statistics}

\begin{enumerate}
\def\labelenumi{(\arabic{enumi})}
\setcounter{enumi}{6}
\tightlist
\item
  Binomial and multinomial count models: beta-binomial models
  \autocite{king1999binomial}, a binomial model for cell-specific counts
  rather than row/column-aggregate counts
  \autocite{wakefield2004ecological}, \emph{R×C} multinomial Dirichlet
  count models \autocite{rosen2001bayesian}, \emph{R×C} models allowing
  for correlation across precincts \autocite{greiner2009rxc}
\item
  Relaxing parametric assumptions on the error term:
  \textcite{imai2008bayesian}
\item
  Geographic adjacency: \textcite{calvo2003local},
  \textcite{anselin2002spatial}
\item
  Local smoothing and geographic subgroup clustering:
  \textcite{chambers2001simple}, \textcite{puig2015ecological}
\item
  Diagnostic tools for Goodman regression: \textcite{gelman2001models}
\item
  Integrating survey data: \textcite{greiner2010exit},
  \textcite{glynn2008alleviating}
\item
  A fast method-of-moments estimator reallocating residuals:
  \textcite{lewis2001understanding}
\item
  Fast approximations for local estimates:
  \textcite{grofman2004ecological}
\item
  Temporal dependency across precincts: \textcite{quinn2004temporal},
  \textcite{lewis2004multiple}
\item
  Using ecological inference output as an independent variable:
  \textcite{herron2003using}
\end{enumerate}

\begin{rrchange}

\emph{Comment}: \textcite{king1997solution} is extensively discussed in
our paper; also see (13). Both (7) and (8) are also discussed in the
main text. (9) and (10) seek to leverage the information that geographic
adjacency provides when the groupings are geographic and their
coordinates are available. Others have thought to improve estimation by
integrating survey data (12). Proposals such as (15) and (16) consider
models and pitfalls for common types of ecological data.

The use of covariates is not the focus for this group of papers. For
example, (11) discuss diagnostics for when regression is not
appropriate, but not covariates. However, some of these propose ways to
incorporate covariates in a particular functional form:
\textcite{ansolabehere1995bias}, \textcite{king1999binomial},
\textcite{wakefield2004ecological}, and \textcite{park2008ecological}.

\end{rrchange}

\subsection*{Partial identification with
bounds}\label{partial-identification-with-bounds}
\addcontentsline{toc}{subsection}{Partial identification with bounds}

\begin{enumerate}
\def\labelenumi{(\arabic{enumi})}
\setcounter{enumi}{16}
\tightlist
\item
  Narrowing Duncan--Davis bounds by assumptions on the functional form
  of contextual effects: \textcite{jiang2020ecological},
  \textcite{manski2018credible}, \textcite{NBERw34285}
\item
  Derived bounds in a wider class of regressions:
  \textcite{cross2002regressions}
\item
  Sampling uncertainty on the bounds: \textcite{fan2016estimation}
\end{enumerate}

\begin{rrchange}

\emph{Comment}: Our article has focused on point estimation. There is a
rich literature on partial identification with bounds in several
statistical traditions as well (17). In economics, ecological inference
is treated as an example of a data integration problem---see (18) and
(19).

\end{rrchange}

\subsection*{In other disciplinary traditions, work distinct from King's
model}\label{in-other-disciplinary-traditions-work-distinct-from-kings-model}
\addcontentsline{toc}{subsection}{In other disciplinary traditions, work
distinct from King's model}

\begin{enumerate}
\def\labelenumi{(\arabic{enumi})}
\setcounter{enumi}{19}
\tightlist
\item
  Ecological inferences with continuous predictor (price) and
  instruments for predictor: \textcite{berry2004differentiated} (BLP);
  flexible functional forms with microdata:
  \textcite{berry2024nonparametric}
\item
  Modeling with individual data and aggregate outcomes:
  \textcite{flaxman2015supported}, \textcite{rosenman2018using},
  \textcite{fishman2024estimating}
\item
  Extensions of Thomsen's regression: \textcite{park2008ecological}
  incorporates covariates, while \textcite{pavia2024ecolrxc} generate
  local estimates and provide software for \emph{R×C} tables
\item
  Linear programming with constraints: \textcite{pavia2024improving}
\end{enumerate}

\begin{rrchange}

\emph{Comment}: The field of industrial organization in economics, as in
(20) is an example of estimating individual parameters with aggregate
data. (21) work in settings where a regression with covariates and
outcome are jointly observed at the individual level (in a sample),
combined with a constraint that their sums add up to a certain count. As
discussed in our paper, optimization approaches such as (23) can be
thought of as passing parametric assumptions to the loss function. The
Thomsen model has been extended in (22) with similar extensions
discussed in our paper.

\end{rrchange}

\section{Proofs of propositions}\label{sec-app-proofs}

\subsection{Relationship between local and global
parameters}\label{relationship-between-local-and-global-parameters}

The local means are connected to the global mean in a similar way as the
individual data are connected to the local parameters.

\[
B_k
:= \frac{1}{N_k} \sum_{i\in I_{k}} Y_i
= \frac{1}{N_k}\sum_{g} \sum_{i\in I_{gk}} Y_i
= \frac{1}{N_k}\sum_{g} \underbrace{N_{gk}\frac{1}{N_{gk}}}_{=1}\sum_{i\in I_{gk}} Y_i
= \frac{\sum_{g} N_{gk}B_{gk}}{\sum_{g} N_{gk}}.
\] The last equality follows from Eq.~\ref{eq-bk} and because
\(N_k=\sum_{g} N_{gk}\).

\subsection{\texorpdfstring{Proof of
Proposition~\ref{prp-id-ccar}}{Proof of Proposition~}}\label{proof-of-prp-id-ccar}

\begin{proof}
When \(\vb Z\) is empty, CAR and CCAR coincide. In this case,
Proposition~\ref{prp-id-ccar} follows immediately from
Proposition~\ref{prp-id}, because as shown in the main text,
Proposition~\ref{prp-id} implies that the true CEF is linear in
\(\bar{\vb X}_g\).
\end{proof}

\subsection{\texorpdfstring{Proof of
Proposition~\ref{prp-id}}{Proof of Proposition~}}\label{proof-of-prp-id}

\begin{proof}
The CAR assumption implies
\(\E[\vb B_g\mid \vb Z_g,\bar{\vb X}_g]=\E[\vb B_g\mid \vb Z_g]\) by the
law of total expectation (integrating out \(N_g\)). Applying this
property once to drop the conditioning on \(\bar X_{gk}=1\) (which, by
the sum-to-1 constraint, fixes \(\bar{\vb X}_g\)) and applying CAR to
condition on \(\bar{\vb X}_g\) and \(N_g\), \[
\begin{aligned}
\E[N_{gk}\E[\bar Y_g\mid \vb Z_g, \bar{X}_{gk}=1]]
&= \E[N_{gk} \E[\vb B_g^\top \bar{\vb X}_g\mid \vb Z_g, \bar{X}_{gk}=1]] \\
&= \E[N_{gk} \E[B_{gk} \mid \vb Z_g, \bar{X}_{gk}=1]] \\
&= \E[N_{gk} \E[B_{gk} \mid \vb Z_g]] \\
&= \E[N_{gk} \E[B_{gk} \mid \vb Z_g, \bar{\vb X}_g, N_g]] \\
&= \E[\E[N_{gk} B_{gk} \mid \vb Z_g, \bar{\vb X}_g, N_g]] \\
&= \E[N_{gk} B_{gk}] = \E[N_{gk}]\beta_k.
\end{aligned}
\] Dividing by \(\E[N_{gk}]\) yields the result.
\end{proof}

\section{Details on Other Models}\label{sec-app-other-models}

\subsection{King's 2×2 model}\label{kings-22-model}

This appendix gathers additional detail on the ecological inference
models surveyed in Section~\ref{sec-est} that is useful for context but
not essential to the main argument. Beyond the discussion in
Section~\ref{sec-est}, a further distinction between King's model and
Goodman's model is in its use of the local and global parameters.
Goodman's regression does not estimate the \(\vb B_g\) and can only
estimate the superpopulation parameter. In King, the global \(\beta\) is
readily calculable as the mean of the estimated truncated Normal
\(\Norm_{[0,1]^2}(\vbg\mu, \Sigma)\) distribution (our \(\vbg\mu\)
corresponds to King's \(\mathfrak{B}\)). Because the form of King's
model and Goodman's regression are the same, they will produce the same
estimates for \(\vbg\beta\) asymptotically.

In the causal inference analogy, the mean of
\(\Norm_{[0,1]^2}(\vbg\mu, \Sigma)\) estimates the superpopulation
\(\vbg\beta\), while the mean of the posterior distribution of
\(\vb B_g\) estimates the finite-sample effect \(\vb B\). However, King
does not advocate using this as the estimate of \(\vbg\beta\). Rather,
because he works in the Bayesian framework, he recommends estimating
\(\vb B\) itself from a weighted mean of the estimated \(\vb B_g\). The
posterior distribution of these local parameters is the original
\(\Norm_{[0,1]^2}(\vbg\mu, \Sigma)\) distribution restricted to the
tomography line defined by the accounting identity,
\(\bar Y_g = B_{g1}\bar X_{g1} + B_{g2}\bar X_{g2}\). Beyond very small
samples, the difference between \(\vb B\) and \(\vbg\beta\) is often
negligible. \textcite{grofman2004ecological} suggest some approximations
to obtain local parameters from a Goodman regression.

\section{Details on Empirical Validations}\label{sec-app-data}

\setcounter{figure}{0}
\setcounter{table}{0}
\renewcommand{\thefigure}{S\arabic{figure}}
\renewcommand{\thetable}{S\arabic{table}}

\subsection{Wallace 1968: Turnout and the Voting Rights
Act}\label{sec-app-wallace-turnout}

The y-axis in Figure~\ref{fig-wal} measures Wallace votes as a share of
the voting-age population (VAP), not as a share of votes cast. This
complicates the interpretation because 1968 was the first general
election following the Voting Rights Act of 1965, and Black voter
registration and turnout in the South lagged that of White voters.
Turnout registration statistics by racial group are not available for
these historical elections, but a 1969 U.S. Census report estimates,
using surveys, that turnout among Black registrants in this election
lagged White voters by approximately 10 points in the South (see
\url{http://bit.ly/4lsvOKu}). By using VAP as the denominator, the
y-axis in Figure~\ref{fig-wal} treats not turning out as an implicit
outcome category, so the figure is in part an ecological inference about
differential turnout rather than vote choice alone. This does not affect
the substantive conclusion---Black voters overwhelmingly did not support
Wallace---but it should be kept in mind when reading the magnitudes.

\subsection{Racial Voting Estimates}\label{racial-voting-estimates}

The voter file we used to create our precinct data was downloaded from
the North Carolina State Board of Elections (NCSBE) public website in
July 2025. We then standardized precinct labels so that it would match
exactly to a precinct in the NCSBE's precinct 2025 shapefiles. We merged
in covariates by obtaining block-group level ACS 2023 estimates of
education, age, and income, projecting counts to the block level
according to total population, and then aggregated the resulting counts
up to the precinct level.

Table~\ref{tbl-nc2-full} compares the estimates of four ecological
inference models in tabular form. See Figure~\ref{fig-nc2} and the main
text for the descriptions of each estimate.

\begin{table}

\caption{\label{tbl-nc2-full}\textbf{Accuracy of EI Methods in
Uncovering Partisanship among Racial Groups, all Outcome Categories.}
Estimates are point estimates with standard errors in parentheses.
Estimates are for the conditional expectation \emph{E(Outcome \textbar{}
Predictor)}. The truth column indicates the voter file ground truth. The
estimates for Democrats are shown in Figure~\ref{fig-nc2}.}

\centering{

\fontsize{10.0pt}{12.0pt}\selectfont
\begin{tabular*}{\linewidth}{@{\extracolsep{\fill}}>{\centering\arraybackslash}p{\dimexpr 112.50pt -2\tabcolsep-1.5\arrayrulewidth}>{\raggedleft\arraybackslash}p{\dimexpr 30.00pt -2\tabcolsep-1.5\arrayrulewidth}cccc}
\toprule
 &  & \multicolumn{4}{c}{{Estimates}} \\ 
\cmidrule(lr){3-6}
Estimand & Truth & Seine & Goodman & Rosen & King \\ 
\midrule\addlinespace[2.5pt]
Pr(Dem | White) & 0.195 & 0.176 (0.008) & 0.175 (0.005) & 0.178 (0.003) & 0.138 (0.002) \\ 
Pr(Dem | Black) & 0.738 & 0.842 (0.031) & 0.848 (0.031) & 0.833 (0.015) & 0.858 (0.005) \\ 
Pr(Dem | Hispanic) & 0.332 & 0.287 (0.042) & 0.081 (0.003) & 0.123 (0.044) & 0.859 (0.011) \\ 
Pr(Ind. | White) & 0.383 & 0.374 (0.013) & 0.364 (0.011) & 0.381 (0.003) & 0.398 (0.001) \\ 
Pr(Ind. | Black) & 0.234 & 0.199 (0.009) & 0.168 (0.006) & 0.153 (0.008) & 0.224 (0.003) \\ 
Pr(Ind. | Hispanic) & 0.488 & 0.638 (0.037) & 0.806 (0.031) & 0.814 (0.077) & 0.275 (0.012) \\ 
Pr(GOP | White) & 0.422 & 0.450 (0.011) & 0.461 (0.014) & 0.441 (0.005) & 0.443 (0.001) \\ 
Pr(GOP | Black) & 0.028 & -0.041 (0.007) & -0.016 (0.002) & 0.014 (0.008) & 0.007 (0.000) \\ 
Pr(GOP | Hispanic) & 0.180 & 0.075 (0.043) & 0.113 (0.005) & 0.063 (0.036) & 0.003 (0.000) \\ 
\bottomrule
\end{tabular*}

}

\end{table}%

Past work has compared different EI methods in the racial polarization
application. \textcite{de2015evidence} evaluates five EI methods on
similar voter file data and finds that \textcite{king1997solution}'s 2×2
and \textcite{rosen2001bayesian}'s count model have lower estimation
error than Goodman's regression, but only explores where and why errors
differ across methods in a cursory fashion.\footnote{\textcite{de2015evidence}
  computes the average precinct-level error in the estimates of
  percentage White, Black, and Hispanic voters registering Democrat. He
  finds that Goodman regression's error is over 68 points, Goodman
  regression with post-hoc truncation has an average error of 25 points,
  and the \textcite{king1997solution} and \textcite{rosen2001bayesian}
  methods both have errors around 15 points. Even the smallest of these
  errors is still substantial. We suspect that estimating the global
  parameter (district or statewide quantities) rather than
  precinct-level local parameters is much less error-prone.
  \textcite{de2015evidence} discusses how differences in errors may be
  explained by the racial composition of the five states he studies
  (373) or the racial composition of the county (377) but does not lay
  out a specific hypothesis.} \textcite{barreto2022estimating} compares
differences between \textcite{king1997solution} and
\textcite{rosen2001bayesian} methods on election results and finds
similar estimates, but does not diagnose which one more correctly
captures the ground truth. \textcite{kuriwaki2024geography} studies the
accuracy of \textcite{rosen2001bayesian}'s method on a Florida voter
file and finds that confidence intervals are too tight.

\begin{table}

\caption{\label{tbl-nc-rosen-sensitivity}\textbf{Count Models' Estimates
are Sensitive to Treatment of Small Groups.} The ``3 x 4'' column shows
estimates from Rosen et al.'s model used in the main text, where the
0.6\% of the population that are registered Libertarian are grouped
together with Independents. The ``4 x 4'' shows the estimates when
Libertarians are treated as a separate outcome category. Standard errors
in parentheses. Estimates change substantively by this specification,
even those for other outcome categories (Democrat and Republican). North
Carolina data.}

\centering{

\fontsize{10.0pt}{12.0pt}\selectfont
\begin{tabular*}{\linewidth}{@{\extracolsep{\fill}}crcc}
\toprule
 &  & \multicolumn{2}{c}{{Estimates, by eiPack specification}} \\ 
\cmidrule(lr){3-4}
Estimand & Truth & 3 x 4 & 4 x 4 \\ 
\midrule\addlinespace[2.5pt]
Pr(GOP | White) & 0.422 & 0.441 (0.005) & 0.354 (0.005) \\ 
Pr(GOP | Black) & 0.028 & 0.014 (0.008) & 0.226 (0.011) \\ 
Pr(GOP | Hispanic) & 0.180 & 0.063 (0.036) & 0.262 (0.005) \\ 
Pr(Dem | White) & 0.195 & 0.178 (0.003) & 0.279 (0.005) \\ 
Pr(Dem | Black) & 0.738 & 0.833 (0.015) & 0.424 (0.014) \\ 
Pr(Dem | Hispanic) & 0.332 & 0.123 (0.044) & 0.359 (0.008) \\ 
\bottomrule
\end{tabular*}

}

\end{table}%

\subsection{Cast Vote Record Data}\label{cast-vote-record-data}

\begin{figure}

\begin{minipage}[c]{0.50\linewidth}

\centering{

\pandocbounded{\includegraphics[keepaspectratio]{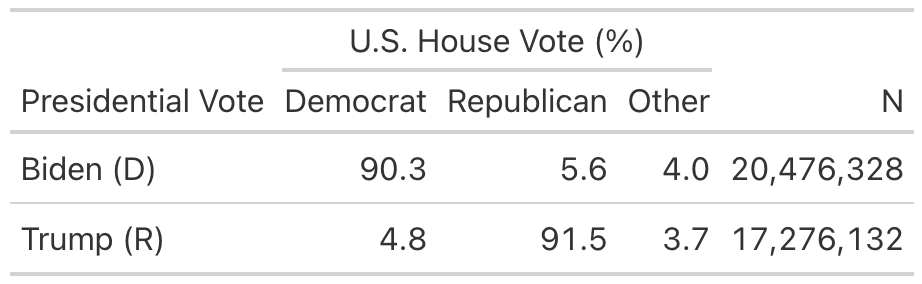}}

}

\subcaption{\label{fig-cvr-all-records}All available records}

\end{minipage}%
\begin{minipage}[c]{0.50\linewidth}

\centering{

\includegraphics[width=0.8\linewidth,height=\textheight,keepaspectratio]{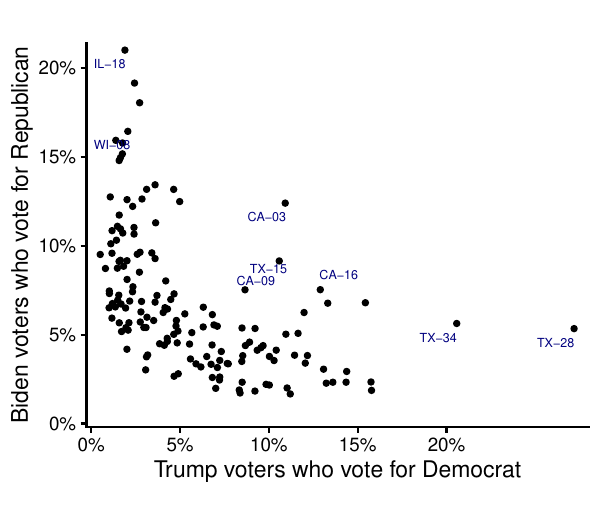}

}

\subcaption{\label{fig-cvr-c}Estimands, by district}

\end{minipage}%

\caption{\label{fig-cvr-1}\textbf{Ticket Splitting Patterns from
Election Data.} The distribution of vote choices in the 2020 election
data from ground truth cast vote records. (a) Across all districts
available in the dataset, approximately 5\% of Biden voters and Trump
voters vote for a different party's candidates in the U.S. House race.
These quantities of interest are off-diagonal entries and cancel out
when aggregated. (b) Quantities of interest, separated for each
congressional district. In typical districts ticket splitting is
asymmetric and benefits one candidate more than their opponent.}

\end{figure}

Figure~\ref{fig-cvr-1} shows an overview of this data. The table in
panel (a) shows row percentages that sum to 100\% in each row. Among the
20 million or so ballots cast for Joseph Biden, the Democratic candidate
for President, 5.6\% voted for a Republican candidate, and 4\% either
skipped the House race (undervote) or voted for a third-party candidate.
In panel (b) we show the level of ticket splitting among Trump voters
(on the horizontal axis) against the level of ticket splitting among
Biden voters in the same district (on the vertical axis). The two
quantities of interest are negatively correlated. Some districts exhibit
substantial ticket splitting for the Republican, and others exhibit
splitting for the Democrat. These two variables cancel out when added to
the entire population in (a).

We limit our analysis to the districts that are contested by a
Democratic and Republican House candidate and whose CVR data contain at
least \(3\) different counties. We impose the latter condition because
the covariates we will use are measured at the county level, so a
district contained wholly within one county has no variation in the
covariate. The median congressional district fragment contains \(177\)
precincts.\footnote{Ballot records are collected county by county. Each
  district's collection of ballots is only a partial set of the
  district's counties. The districts of the final data are in AZ, CA,
  CO, FL, GA, IL, MD, MI, NJ, NV, OH, OR, TX, and WI.}

\textcite{burden1998new} and \textcite{burden2009americans} were the
first to use ecological inference methods, applying King's
(\autocite*{king1997solution}) algorithm on congressional district level
aggregate data. \textcite{tam2004limits} critiqued the uninformativeness
of aggregate data in the case of ticket splitting. Neither work had
access to ground truth levels of ticket splitting. Other countries such
as Austria and New Zealand reported vote results in a way that allowed
ticket splitting rates to be computed from ballots
\autocite{klima2016estimation,pavia2024improving}. By the 2010s, as some
U.S. states started to release cast vote records,
\textcite{park2014ecological} evaluated King's EI model and Thomsen's
nonlinear regression model in ten counties. They did not test other
models or include covariates.

\subsection{Sensitivity Analysis}\label{sensitivity-analysis}

Figure~\ref{fig-sens} estimates sensitivity bounds with the following
specifications.

\begin{itemize}
\tightlist
\item
  Wallace data: We expand the example from Figure~\ref{fig-wal} to
  eleven former Confederate states. We include as covariates a
  categorical variable for state, proportion of the population in farm
  households, proportion of the population with elementary, high school,
  and college education, proportion of the population in households in
  one of four income brackets. All data is at the county level.
\item
  North Carolina data: We use the same data and covariates in
  Figure~\ref{fig-nc2}.
\item
  Cast vote record data: We use the same data and covariates as used in
  Figure~\ref{fig-cvr-2}.
\end{itemize}

For all three datasets, we compute a DML estimate. For computational
feasibility we enter the covariates linearly, without interactions or
spline transformation.

\newpage{}

\printbibliography[title=References]
\end{refsection}

% Intentionally empty.
% This format prints separate main-text and appendix bibliographies manually with
% biblatex refsections, so Quarto's default bibliography partial must not emit a
% third bibliography at the end of the document.
% Intentionally empty.
% The paper prints separate main-text and appendix bibliographies manually with
% biblatex refsections; Quarto's default after-body partial would add a third
% automatic bibliography outside those refsections.
\end{document}